 \documentclass[authoryear,preprint,12pt]{elsarticle}
\usepackage{colortbl}

\usepackage{amssymb}
\usepackage{multirow}
\usepackage{booktabs,subcaption,amsfonts,dcolumn}
\newcolumntype{d}[1]{D..{#1}}

\usepackage{algorithmic}
\usepackage{graphicx}
\usepackage{amsmath}

\usepackage{amssymb}
\usepackage{url}
\usepackage{geometry}
\usepackage{color}
\usepackage{booktabs}
\usepackage{dsfont}
\usepackage[ruled,vlined]{algorithm2e}
\usepackage{hyperref}
\usepackage[x11names]{xcolor}

\definecolor{RED}{rgb}{1,0,0}
\definecolor{ORANGE}{rgb}{1,0.5,0}
\definecolor{BLUE}{rgb}{0,0,1}

\newcommand{\indep}{\perp \!\!\! \perp}
\usepackage{pgf, tikz} 
\usetikzlibrary{arrows,automata,fit}
\usetikzlibrary{arrows.meta}
\usetikzlibrary{shapes}
\usetikzlibrary{arrows,shapes.geometric,graphs,graphs.standard,positioning}

\newcommand{\oo}{7.1}
\newcommand{\pp}{8.45}

\newcommand{\rr}{3.6}
\newcommand{\bb}{4.6}

\newcommand{\xx}{2.1}
\newcommand{\zz}{1.2}
\newcommand{\yy}{2.4}

\newcommand{\xxx}{4.3}
\newcommand{\zzz}{1.05}
\newcommand{\yyy}{6.5}

\usepackage{makecell}
\usepackage{bm}
\usepackage{lscape}

\newcommand{\sag}[2]{\tikz{\node[shape=circle,draw,inner sep=1pt,minimum width = 0.6cm, fill=#1]{$X_{#2}$};}} 
\usepackage{verbatim}

\usepackage{algorithmic}
\ifpdf
\DeclareGraphicsExtensions{.eps,.pdf,.png,.jpg}
\else
\DeclareGraphicsExtensions{.eps}
\fi

\journal{Socio-Economic Planning Sciences}

\begin{document}

\begin{frontmatter}



\title{Robust learning of staged tree models: A case study in evaluating transport services}


\author[ie]{Manuele Leonelli}
\author[gherardo]{Gherardo Varando}

\affiliation[ie]{organization={School of Science and Technology, IE University},
            city={Madrid},
            country={Spain}}
            
\affiliation[gherardo]{organization={Image Processing Laboratory, Universitat de Valencia},
            city={Valencia},
            country={Spain}}

\begin{abstract}
Staged trees are a relatively recent class of probabilistic graphical models that extend Bayesian networks to formally and graphically account for non-symmetric patterns of dependence. Machine learning algorithms to learn them from data have been implemented in various pieces of software. However, to date, methods to assess the robustness and validity of the learned, non-symmetric relationships are not available. Here, we introduce validation techniques tailored to staged tree models based on non-parametric bootstrap resampling methods and investigate their use in practical applications. In particular, we focus on the evaluation of transport services using large-scale survey data. In these types of applications, data from heterogeneous sources must be collated together. Staged trees provide a natural framework for this integration of data and its analysis. For the thorough evaluation of transport services, we further implement novel what-if sensitivity analyses for staged trees and their visualization using software.

\end{abstract}








\end{frontmatter}


\section{Introduction}
\label{sec:introduction}
Probabilistic graphical models (PGMs) represent the interactions between variables of interest using a graph. They decompose a joint probability mass function (pmf) into the product of local, smaller dimensional conditional pmfs, which are determined by the underlying graph. Because of this decomposition, they are tailored for the integration of heterogeneous data and expert sources, since each source can coherently and independently inform different local distributions \citep{johnson2012integrated,leonelli2020coherent,marcot2019advances}. Bayesian networks (BNs) are undoubtedly the most commonly used PGM, describing the underlying dependence structure through a directed acyclic graph (DAG) \citep{pearl1988probabilistic}. 


One of the limitations of BNs is that they can formally encode only symmetric conditional independences between variables, determined by the famous d-separation criterion \citep[e.g.][]{pearl2009causality}. However, several studies have now shown that more complex patterns of dependence are required to faithfully describe the information stored in collected data \citep{eggeling2019algorithms,jaeger2006learning,pensar2015labeled,talvitie2019learning}. The most classical non-symmetric independence is context-specific \citep{Boutilier1996}, where independence holds only for a specific subset of values of the conditioning variables. More generic types of non-symmetric independence have been defined and studied in the context of PGMs \citep{pensar2016role}.

Although it was recognized early on the need for PGMs embedding non-symmetric independence \citep{chickering1997bayesian,friedman1996learning}, the development of such models has been limited. Staged trees \citep{collazo2018chain,smith2008conditional} are a class of non-symmetric PGMs that embed flexible types of non-symmetric independence by coloring the vertices of a probability tree. There is now an extensive toolkit of methodologies for analyzing and learning staged trees from data, as well as user-friendly software for their use, including multiple structural learning algorithms for the identification of non-symmetric independence from data \citep{Carli2022,walley2023cegpy}.

Much attention has been devoted to the development of robust learning algorithms for BNs, often based on the use of sampling techniques. \citet{friedman1999data} and \citet{caravagna2021learning} investigated the use of data bootstrapping to learn the underlying DAG: a graph is learned for each bootstrap replication and edges in the final BN are chosen depending on how often they appeared \citep{scutari}. \citet{friedman2003being} introduced Bayesian MCMC approaches of learning, whose use is becoming increasingly popular \citep{castelletti2021equivalence,goudie2016gibbs,kuipers2017partition,kuipers2022efficient,viinikka2020layering}. \citet{cugnata2016bayesian} discussed the averaging of BNs learned with different structural learning algorithms: a technique used in several practical applications \citep[e.g.][]{ceriani2020multidimensional,di2017monitoring,mandhani2021establishing}. Cross-validation is often further used to select the best BN structure with the aim of avoiding overfitting \citep[e.g.][]{liew2022short}.

Despite the widespread use of such techniques for BNs, their use has not been discussed for staged trees. Structural learning algorithms are usually evaluated over the complete dataset, thus risking overfitting and identifying relationships that do not actually represent the underlying dependence structure. In this direction, \citet{strong2022bayesian} introduced a Bayesian model averaging approach to jointly consider high-scoring staged trees for more robust modeling.

This paper introduces robust routines for learning non-symmetric conditional independence via staged trees, discussing their methodological and software implementation. Methodologically, we first discuss the robust choice of a variables' ordering using bootstrap. Once such an order is identified, bootstrap and cross-validation are jointly used to identify the best staged tree model from data. Software-wise, we take advantage of the capabilities of the \texttt{stagedtrees} R package \citep{Carli2022}. 

Our novel, robust learning methods are illustrated to untangle complex dependence patterns in survey data for the evaluation of transport services. BNs have been used  for this task in metro \citep{diez2018bayesian,hua2021bayesian,mandhani2020interrelationships,mandhani2021establishing,xu2020improving}, railways \citep{perucca2014travellers}, airlines \citep{cugnata2016bayesian}, airport check-in \citep{di2017monitoring}, and air and high-speed rail intermodal \citep{yang2022exploring} services. The comprehensive evaluation of transport services requires the integration of heterogeneous data sources, often collected at different spatial or temporal resolutions. This integration is exemplified in our second data application below about the quality of railway services in the European Union.

A thorough assessment of the factors related to the evaluation of transport services requires the use of sensitivity techniques \citep[e.g.][]{borgonovo2023sensitivity}. In the context of BNs, what-if analyses are widely used, where specific vertices are assumed to be observed and the effect of these observations on output of interest is measured \citep[e.g.][]{cugnata2016bayesian}. In our applications below, we showcase a novel visual way to perform what-if analyses in staged trees.

The paper is structured as follows. Section \ref{sec2} deals with PGMs, reviewing BNs and staged trees. Section \ref{sec3}, after reviewing structural learning algorithms for staged trees, introduces robust routines based on the bootstrap. Section \ref{sec4} showcases these methods over two data applications: the first in airline passengers' satisfaction, the second based on a large-scale European survey on railways satisfaction. The paper concludes with a discussion.

\section{Non-symmetric probabilistic graphical models}
\label{sec2}

Probabilistic graphical models (PGMs) \citep{koller2009probabilistic} are a popular class of statistical models that use various graphical representations to visually depict the dependence structure between variables of interest. In this section, we review two instances of PGMs: the most common Bayesian network (BN) model and the staged tree model.

\subsection{Bayesian networks and conditional independence}

First introduced by \citet{pearl1988probabilistic}, BNs are a class of PGMs that use directed acyclic graphs (DAGs) to formally encode independence information. They are now the gold standard for representing causal information and discovering it from observational data \citep{glymour2019review}. The variables of interest are the vertices of the DAG. The edges denote probabilistic direct dependence and, in some cases, causal relationships. Although BNs can be used with continuous and mixed variables, henceforth, and as common in practice, we focus on the categorical case. Let $X=(X_1,\dots,X_p)$ be a categorical random vector and $x=(x_1,\dots,x_p)$ an instantiation of this vector. 


\begin{figure}
\centering
\begin{tikzpicture}
\node (1) at (0*\xx,0.5*\yy){\sag{white}{1}};
\node (2) at (1.5*\xx,0*\yy){\sag{white}{2}};
\node (3) at (1.5*\xx,1*\yy){\sag{white}{3}};
\node (4) at (3*\xx,0.5*\yy){\sag{white}{4}};
\draw[->, line width = 1.1pt] (1) -- (2);
\draw[->, line width = 1.1pt] (1) -- (3);
\draw[->, line width = 1.1pt] (3) -- (4);
\draw[->, line width = 1.1pt] (2) -- (4);
\end{tikzpicture}
\caption{An example of a DAG over four random variables $X_1,X_2,X_3,X_4$.
\label{fig:bn1}}
\end{figure}
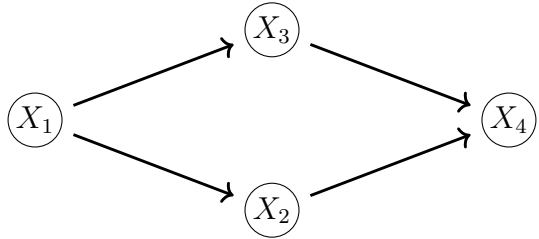

One of the main features of BNs is that they decompose the joint probability mass function (pmf) of $X$, $P(x)$, into a product of smaller-dimensional conditional pmfs. The form of these conditional probabilities is defined by the DAG $G$. Let $X_{\Pi_i}$ be the parents of $X_i$ in $G$. The joint pmf of a BN with DAG $G$ can be written as
\begin{equation}
\label{eq:factorization}
P(x)=\prod_{i=1}^pP(x_i|x_{\Pi_i}),
\end{equation}
where $\Pi_i$ could be empty (by construction, there must be at least one vertex with no parents in a DAG). For instance, the DAG in Figure \ref{fig:bn1} induces the factorization
\begin{equation}
    P(x)=P(x_4|x_3,x_2)P(x_3|x_2)P(x_2|x_1)P(x_1).
\end{equation}
The great benefit associated with the decomposition in Equation \ref{eq:factorization} is that the number of parameters defining the model can decrease dramatically. In our example, if variables are assumed to be binary, nine parameters are required to fully characterize the BN, in contrast to fifteen generally required to define the pmf of four binary variables ($2^4-1$, since probabilities must sum up to one).

The lack of edges in the DAG of a BN formally encodes symmetric \textit{conditional independence} \citep{dawid1979conditional}. We say that $X_i$ is conditionally independent of $X_j$ given $X_k$, and write $X_i\indep X_j|X_k$, if and only if
\begin{equation}
\label{eq:ci}
    p(x_i|x_j,x_k) = p(x_i|x_k),
\end{equation}
for all possible values $x_i,x_j,x_k$. The factorization of the pmf in Equation \ref{eq:factorization} is associated with a set of conditional independences, usually referred to as \textit{ordered Markov property} \citep[e.g.][]{koller2009probabilistic}: each variable is conditionally independent of its non-descendants given its parents. For our example DAG in Figure \ref{fig:bn1} the lack of the edge $(X_1,X_4)$ is associated with the independence $X_4\indep X_1|X_2,X_3$, while the missing $(X_2,X_3)$ represents $X_3\indep X_2 | X_1$. However, the conditional independences from the ordered Markov property are not the only ones associated with a BN. The DAG can be investigated to check if generic conditional independences hold in the model via the so-called \textit{d-separation} criterion \citep[see e.g.][for details]{pearl2009causality}.

\begin{table}
\begin{center}
\scalebox{0.83}{
\begin{tabular}{cccc}
\multicolumn{4}{c}{\textit{Country}}\\
    SE & WE & EE & NE  \\
     \midrule
     0.15&0.25 & 0.35 & 0.25 
\end{tabular}
}
\end{center}

\vspace{-0.4cm}

\begin{center}
\scalebox{0.83}{
\begin{tabular}{c|cc}
&\multicolumn{2}{c}{\textit{Length}}\\
\textit{Country}  &   Low& High  \\
     \midrule
     SE & 0.6 & 0.4 \\
     WE & 0.3 & 0.7 \\
     EE & 0.6 & 0.4 \\
     NE & 0.3 & 0.7
\end{tabular}
\,\,\,
\begin{tabular}{c|cc}
&\multicolumn{2}{c}{\textit{Income}}\\
\textit{Country}  &   Low& High  \\
     \midrule
     SE & 0.5 & 0.5 \\
     WE & 0.2 & 0.8 \\
     EE & 0.7 & 0.3 \\
     NE & 0.2 & 0.8
\end{tabular}
\,\,\,
\begin{tabular}{cc|ccc}
&&\multicolumn{3}{c}{\textit{Satisfaction}}\\
\textit{Length} &\textit{Income}  &   Low& Medium &High  \\
     \midrule
    High & High & 0.1 & 0.4 &0.5 \\
    High &Low& 0.3 & 0.4 & 0.3\\
     Low & High& 0.5 & 0.3&0.2 \\
     Low& Low & 0.5 & 0.3&0.2
\end{tabular}
}
\end{center}
\vspace{-0.4cm}
\caption{CPTs associated to the BN in Figure \ref{fig:bn1}. \label{table:1}}
\end{table}

To better illustrate BNs, consider the following simplified scenario of the much more complex application on railway travel satisfaction we develop in Section \ref{sec:rail}. We are interested in assessing how the length of national railways, average national income, and the country in which the travel took place affect the satisfaction of railway travelers. For simplicity, we grouped European countries into four regions: southern Europe (SE), western Europe (WE), eastern Europe (EE), and northern Europe (NE). We make the assumption that conditionally on the length of the railway and the national income, knowing the region of the traveler is irrelevant to predicting satisfaction. Furthermore, we assume that conditionally on the region, the length of the railway does not provide any information to predict the average income. This situation can be depicted by the DAG in Figure \ref{fig:bn1} with $X_1=$Country, $X_2=$Length, $X_3=$Income, and $X_4$=Satisfaction. The definition of the BN is completed by the specification of the conditional probabilities of the model. In the categorical case, these are most commonly referred to as \textit{conditional probability tables (CPTs)}. These tables store the conditional probability of a variable $X_i$ for every possible combination of its parents $X_{\Pi_i}$. The CPTs for our railway travel satisfaction example are reported in Table \ref{table:1}, assuming Length and Income are categorized into Low and High, while Satisfaction into Low, Medium, and High. Notice that the CPTs automatically embed the conditional independences of the model: for instance, the CPT of Satisfaction states that, conditionally on each combination of Length and Income, its probability distribution is the same for every European region. The probability of any event can then be computed from the CPTs: the probability of a southern European traveler from a high-income country with a low railway track length being highly satisfied can be computed using Equation \ref{eq:factorization} as $0.15\cdot 0.5\cdot 0.6\cdot 0.2=0.009$.

As apparent from the previous discussion, the definition of a BN model consists of two steps: the definition of a DAG $G$ establishing the dependence structure between the variables; and the CPTs storing the conditional probabilities of variables given their parents in $G$. Although these two steps can be performed via expert elicitation \citep{renooij2001probability,werner2017expert,wilkerson2021customized,zhang2016expert}, our focus here is in the case both are learned from observational data using machine learning algorithms. Data-driven algorithms to learn the DAG $G$  are usually referred to as \textit{structural learning algorithms} \citep[see][for an overview]{kitson2023survey,scanagatta2019survey,scutari2019learns}. Three classes of algorithms are common: \textit{constraint-based} algorithms \citep[e.g. the PC algorithm][]{spirtes}, which use conditional independence tests; \textit{score-based} algorithms \citep[e.g. the tabu algorithm][]{russel}, which use goodness-of-fit scores as objective functions to maximize; and
\textit{hybrid} algorithms \citep[e.g. the MMHC algorithm][]{tsamardinos2006max} that combine both approaches. Given a DAG $G$, maximum likelihood or Bayesian approaches can be used to learn the CPT tables. 

BNs provide an efficient platform to answer \textit{inferential queries}, meaning computing (conditional) probabilities of interest from the model. Although this is, in general, an NP-hard problem, algorithms that take advantage of the underlying DAG structure have been defined \citep[e.g.][]{koller2009probabilistic}. In applied analyses, these types of query are usually called \textit{what-if analysis}, which are used to identify the most important factors affecting an output of interest. 

\subsection{Non-symmetric conditional independence}

A BN model can only formally encode symmetric types of conditional independence corresponding to equalities between probabilities as in Equation (\ref{eq:ci}). Those associated with the ordered Markov condition are by construction imposed by the CPTs of the BN since probabilities are defined conditionally on the parent variables only. However, often, the CPTs of a BN include additional equalities between its entries, which cannot be inferred by simply looking at the underlying DAG.

As a first example, consider the CPT for Satisfaction in Table \ref{table:1}. It can be seen that the probability distribution of Satisfaction is the same in the cases Length=Low, Income=High and Length=Low, Income=Low. In other words, conditionally, on a low railway length, income is irrelevant to predicting satisfaction. So this is a conditional independence that holds for only a specific value of the conditioning variable, Length=Low, and not for the other (Length=High). This is usually referred to as a \textit{context-specific} independence \citep{Boutilier1996}. More formally, we say that $X_i$ is context-specific independent of $X_j$ in the context $x_k$ if
\begin{equation}
\label{eq:csi}
p(x_i|x_j,x_k)=p(x_i|x_k)
\end{equation}
for all possible values $x_i,x_j$. Notice that for another value of the variable $X_k$, say $x_k'$, we would have that $p(x_i|x_j,x_k')\neq p(x_i|x_k')$.

The CPTs in Table \ref{table:1} include further equalities between their rows. For instance, the probability distribution of Length is the same for Country=SE,EE, as well as for Country=WE,NE. This means e.g. that the probability of Length=High is the same in Northern and Western Europe. Similarly, there are equalities in the CPT of Income.  These types of equalities have been referred to as \textit{partial conditional independence} \citep{pensar2016role}. More formally, we say that $X_i$ is partially conditionally independent of $X_j$ in the domain $\{x_j^{1},\dots,x_j^{l}\}$ given context $x_k$ if 
\begin{equation}
\label{eq:pci}
P(x_i|x_j^{a},x_k)=P(x_i|x_j^{b},x_k),
\end{equation}
for all values $x_i$ and every pair $x_j^{a},x_j^{b}$ in the domain $\{x_j^{1},\dots,x_j^{l}\}$. Notice that in our example there was no conditioning context $x_k$: this case is sometimes called marginal partial independence. If the domain $\{x_j^{1},\dots,x_j^{l}\}$ includes all possible values $x_j$, then Equations (\ref{eq:csi}) and (\ref{eq:pci}) coincide. Notice that variables must take more than two values for a non-trivial partial conditional independence to hold.

A final type of independence is the so-called \textit{local independence} \citep{chickering1997bayesian}, which simply states that some conditional probability distributions are
identical but no discernable patterns as in Equations (\ref{eq:csi}) and (\ref{eq:pci}) can be detected. Let $X_A$ be a vector not including $X_i$ and $x_A$ and $x_A'$ two instantiations of $X_A$. A local independence is an equality of the form
\begin{equation}
    P(x_i|x_A)=P(x_i|x_A'),
\end{equation}
for all values $x_i$. An illustration of local independence is given in Table \ref{table:2}. The conditional probability distribution of Satisfaction is the same for the case Length=Low, Income=Low and Length=High, Income=High.

\begin{table}[]
    \centering
    \scalebox{0.83}{
\begin{tabular}{cc|ccc}
&&\multicolumn{3}{c}{\textit{Satisfaction}}\\
\textit{Length} &\textit{Income}  &   Low& Medium &High  \\
     \midrule
    High & High & 0.5 & 0.3 &0.2 \\
    High &Low& 0.3 & 0.4 & 0.3\\
     Low & High& 0.7 & 0.2&0.1 \\
     Low& Low & 0.5 & 0.3&0.2
\end{tabular}
}
    \caption{Example of a CPT including a local independence.}
    \label{table:2}
\end{table}

These more generic constraints between probabilities are usually referred to as \textit{non-symmetric conditional independence}. As noticed, BNs are not able to graphically visualize this additional information often included in their CPTs. For this reason, extensions of BNs that can graphically visualize non-symmetric patterns of dependence have been proposed
\citep[e.g.][]{eggeling2019algorithms,jaeger2006learning,pensar2015labeled,talvitie2019learning}. In this paper, we focus on one specific extension of BNs called \textit{staged tree model}.

\subsection{Staged trees}
\label{sec:staged}

Differently to BNs, whose graphical representation is a DAG, staged trees \citep{collazo2018chain,smith2008conditional} are PGMs that visualize conditional independence by means of a colored tree. A \textit{probability tree} is associated with the vector of interest $X=(X_1,\dots,X_p)$, denoting its sample space and probabilities. A probability tree is a rooted directed tree where each edge reports a (conditional) probability. The sum of the probabilities of edges emanating from the same non-leaf vertex must sum up to one and the product of the probabilities of edges along a root-to-leaf path is equal to the probability of the associated atomic event.

\begin{figure}
    \centering
    \scalebox{0.38}{
    \begin{tikzpicture}[auto, scale=10,
	Country_NA/.style={circle,inner sep=0.5mm,minimum size=1.6cm,draw,black,very thick,text=black},
	Length_1/.style={circle,inner sep=0.5mm,minimum size=1.6cm,draw,black,very thick,fill=Cyan1,text=black},
	Length_2/.style={circle,inner sep=0.5mm,minimum size=1.6cm,draw,black,very thick,fill=Firebrick1,text=black},
	Income_1/.style={circle,inner sep=0.5mm,minimum size=1.6cm,draw,black,very thick,fill=Orchid1,text=black},
	Income_2/.style={circle,inner sep=0.5mm,minimum size=1.6cm,draw,black,very thick,fill=Yellow1,text=black},
	Income_3/.style={circle,inner sep=0.5mm,minimum size=1.6cm,draw,black,very thick,fill=Green1,text=black},
	Satisfaction_1/.style={circle,inner sep=0.5mm,minimum size=1.6cm,draw,black,very thick,fill=Aquamarine1,text=black},
	Satisfaction_2/.style={circle,inner sep=0.5mm,minimum size=1.6cm,draw,black,very thick,fill=LightPink1,text=black},
	Satisfaction_3/.style={circle,inner sep=0.5mm,minimum size=1.6cm,draw,black,very thick,fill=Snow2,text=black},
leaf/.style={circle,inner sep=0.5mm,minimum size=0.4cm,draw,very thick,black,fill=gray,text=black}]

	\node [Country_NA] (root) at (0.000000*\rr, 0.500000*\bb)	{\Huge{$v_0$}};
	\node [Length_1] (root-EE) at (0.2500000*\rr, 0.117021*\bb)	{\Huge{$v_1$}};
	\node [Length_2] (root-NE) at (0.2500000*\rr, 0.372340*\bb)	{\Huge{$v_2$}};
	\node [Length_1] (root-SE) at (0.250000*\rr, 0.627660*\bb)	{\Huge{$v_3$}};
	\node [Length_2] (root-WE) at (0.250000*\rr, 0.882979*\bb)	{\Huge{$v_4$}};
	\node [Income_1] (root-EE-Low) at (0.500000*\rr, 0.053191*\bb)	{\Huge{$v_5$}};
	\node [Income_1] (root-EE-High) at (0.500000*\rr, 0.180851*\bb)	{\Huge{$v_6$}};
	\node [Income_2] (root-NE-Low) at (0.500000*\rr, 0.308511*\bb)	{\Huge{$v_7$}};
	\node [Income_2] (root-NE-High) at (0.500000*\rr, 0.436170*\bb)	{\Huge{$v_8$}};
	\node [Income_3] (root-SE-Low) at (0.500000*\rr, 0.563830*\bb)	{\Huge{$v_9$}};
	\node [Income_3] (root-SE-High) at (0.500000*\rr, 0.691489*\bb)	{\Huge{$v_{10}$}};
	\node [Income_2] (root-WE-Low) at (0.500000*\rr, 0.819149*\bb)	{\Huge{$v_{11}$}};
	\node [Income_2] (root-WE-High) at (0.500000*\rr, 0.946809*\bb)	{\Huge{$v_{12}$}};
	\node [Satisfaction_1] (root-EE-Low-Low) at (0.750000*\rr, 0.021277*\bb)	{\Huge{$v_{13}$}};
	\node [Satisfaction_1] (root-EE-Low-High) at (0.750000*\rr, 0.085106*\bb)	{\Huge{$v_{14}$}};
	\node [Satisfaction_2] (root-EE-High-Low) at (0.750000*\rr, 0.148936*\bb)	{\Huge{$v_{15}$}};
	\node [Satisfaction_3] (root-EE-High-High) at (0.750000*\rr, 0.212766*\bb)	{\Huge{$v_{16}$}};
	\node [Satisfaction_1] (root-NE-Low-Low) at (0.750000*\rr, 0.276596*\bb)	{\Huge{$v_{17}$}};
	\node [Satisfaction_1] (root-NE-Low-High) at (0.750000*\rr, 0.340426*\bb)	{\Huge{$v_{18}$}};
	\node [Satisfaction_2] (root-NE-High-Low) at (0.750000*\rr, 0.404255*\bb)	{\Huge{$v_{19}$}};
	\node [Satisfaction_3] (root-NE-High-High) at (0.750000*\rr, 0.468085*\bb)	{\Huge{$v_{20}$}};
	\node [Satisfaction_1] (root-SE-Low-Low) at (0.750000*\rr, 0.531915*\bb)	{\Huge{$v_{21}$}};
	\node [Satisfaction_1] (root-SE-Low-High) at (0.750000*\rr, 0.595745*\bb)	{\Huge{$v_{22}$}};
	\node [Satisfaction_2] (root-SE-High-Low) at (0.750000*\rr, 0.659574*\bb)	{\Huge{$v_{23}$}};
	\node [Satisfaction_3] (root-SE-High-High) at (0.750000*\rr, 0.723404*\bb)	{\Huge{$v_{24}$}};
	\node [Satisfaction_1] (root-WE-Low-Low) at (0.750000*\rr, 0.787234*\bb)	{\Huge{$v_{25}$}};
	\node [Satisfaction_1] (root-WE-Low-High) at (0.750000*\rr, 0.851064*\bb)	{\Huge{$v_{26}$}};
	\node [Satisfaction_2] (root-WE-High-Low) at (0.750000*\rr, 0.914894*\bb)	{\Huge{$v_{27}$}};
	\node [Satisfaction_3] (root-WE-High-High) at (0.750000*\rr, 0.978723*\bb)	{\Huge{$v_{28}$}};
	\node [leaf] (root-EE-Low-Low-High) at (1.000000*\rr, 0.000000*\bb)	{};
	\node [leaf] (root-EE-Low-Low-Medium) at (1.000000*\rr, 0.021277*\bb)	{};
	\node [leaf] (root-EE-Low-Low-Low) at (1.000000*\rr, 0.042553*\bb)	{};
	\node [leaf] (root-EE-Low-High-High) at (1.000000*\rr, 0.063830*\bb)	{};
	\node [leaf] (root-EE-Low-High-Medium) at (1.000000*\rr, 0.085106*\bb)	{};
	\node [leaf] (root-EE-Low-High-Low) at (1.000000*\rr, 0.106383*\bb)	{};
	\node [leaf] (root-EE-High-Low-High) at (1.000000*\rr, 0.127660*\bb)	{};
	\node [leaf] (root-EE-High-Low-Medium) at (1.000000*\rr, 0.148936*\bb)	{};
	\node [leaf] (root-EE-High-Low-Low) at (1.000000*\rr, 0.170213*\bb)	{};
	\node [leaf] (root-EE-High-High-High) at (1.000000*\rr, 0.191489*\bb)	{};
	\node [leaf] (root-EE-High-High-Medium) at (1.000000*\rr, 0.212766*\bb)	{};
	\node [leaf] (root-EE-High-High-Low) at (1.000000*\rr, 0.234043*\bb)	{};
	\node [leaf] (root-NE-Low-Low-High) at (1.000000*\rr, 0.255319*\bb)	{};
	\node [leaf] (root-NE-Low-Low-Medium) at (1.000000*\rr, 0.276596*\bb)	{};
	\node [leaf] (root-NE-Low-Low-Low) at (1.000000*\rr, 0.297872*\bb)	{};
	\node [leaf] (root-NE-Low-High-High) at (1.000000*\rr, 0.319149*\bb)	{};
	\node [leaf] (root-NE-Low-High-Medium) at (1.000000*\rr, 0.340426*\bb)	{};
	\node [leaf] (root-NE-Low-High-Low) at (1.000000*\rr, 0.361702*\bb)	{};
	\node [leaf] (root-NE-High-Low-High) at (1.000000*\rr, 0.382979*\bb)	{};
	\node [leaf] (root-NE-High-Low-Medium) at (1.000000*\rr, 0.404255*\bb)	{};
	\node [leaf] (root-NE-High-Low-Low) at (1.000000*\rr, 0.425532*\bb)	{};
	\node [leaf] (root-NE-High-High-High) at (1.000000*\rr, 0.446809*\bb)	{};
	\node [leaf] (root-NE-High-High-Medium) at (1.000000*\rr, 0.468085*\bb)	{};
	\node [leaf] (root-NE-High-High-Low) at (1.000000*\rr, 0.489362*\bb)	{};
	\node [leaf] (root-SE-Low-Low-High) at (1.000000*\rr, 0.510638*\bb)	{};
	\node [leaf] (root-SE-Low-Low-Medium) at (1.000000*\rr, 0.531915*\bb)	{};
	\node [leaf] (root-SE-Low-Low-Low) at (1.000000*\rr, 0.553191*\bb)	{};
	\node [leaf] (root-SE-Low-High-High) at (1.000000*\rr, 0.574468*\bb)	{};
	\node [leaf] (root-SE-Low-High-Medium) at (1.000000*\rr, 0.595745*\bb)	{};
	\node [leaf] (root-SE-Low-High-Low) at (1.000000*\rr, 0.617021*\bb)	{};
	\node [leaf] (root-SE-High-Low-High) at (1.000000*\rr, 0.638298*\bb)	{};
	\node [leaf] (root-SE-High-Low-Medium) at (1.000000*\rr, 0.659574*\bb)	{};
	\node [leaf] (root-SE-High-Low-Low) at (1.000000*\rr, 0.680851*\bb)	{};
	\node [leaf] (root-SE-High-High-High) at (1.000000*\rr, 0.702128*\bb)	{};
	\node [leaf] (root-SE-High-High-Medium) at (1.000000*\rr, 0.723404*\bb)	{};
	\node [leaf] (root-SE-High-High-Low) at (1.000000*\rr, 0.744681*\bb)	{};
	\node [leaf] (root-WE-Low-Low-High) at (1.000000*\rr, 0.765957*\bb)	{};
	\node [leaf] (root-WE-Low-Low-Medium) at (1.000000*\rr, 0.787234*\bb)	{};
	\node [leaf] (root-WE-Low-Low-Low) at (1.000000*\rr, 0.808511*\bb)	{};
	\node [leaf] (root-WE-Low-High-High) at (1.000000*\rr, 0.829787*\bb)	{};
	\node [leaf] (root-WE-Low-High-Medium) at (1.000000*\rr, 0.851064*\bb)	{};
	\node [leaf] (root-WE-Low-High-Low) at (1.000000*\rr, 0.872340*\bb)	{};
	\node [leaf] (root-WE-High-Low-High) at (1.000000*\rr, 0.893617*\bb)	{};
	\node [leaf] (root-WE-High-Low-Medium) at (1.000000*\rr, 0.914894*\bb)	{};
	\node [leaf] (root-WE-High-Low-Low) at (1.000000*\rr, 0.936170*\bb)	{};
	\node [leaf] (root-WE-High-High-High) at (1.000000*\rr, 0.957447*\bb)	{};
	\node [leaf] (root-WE-High-High-Medium) at (1.000000*\rr, 0.978723*\bb)	{};
	\node [leaf] (root-WE-High-High-Low) at (1.000000*\rr, 1.000000*\bb)	{};
	\draw[-{Latex[width=3mm,length=3mm]}] (root) -- node [sloped,pos=0.8]{\LARGE{EE}} (root-EE);
	\draw[-{Latex[width=3mm,length=3mm]}] (root) -- node [sloped,pos=0.8]{\LARGE{NE}} (root-NE);
	\draw[-{Latex[width=3mm,length=3mm]}] (root) -- node [sloped,pos=0.8]{\LARGE{SE}} (root-SE);
	\draw[-{Latex[width=3mm,length=3mm]}] (root) -- node [sloped,pos=0.8]{\LARGE{WE}} (root-WE);
	\draw[-{Latex[width=3mm,length=3mm]}] (root-EE) -- node [sloped,pos=0.8]{\LARGE{Low}} (root-EE-Low);
	\draw[-{Latex[width=3mm,length=3mm]}] (root-EE) -- node [sloped,pos=0.8]{\LARGE{High}} (root-EE-High);
	\draw[-{Latex[width=3mm,length=3mm]}] (root-NE) -- node [sloped,pos=0.8]{\LARGE{Low}}  (root-NE-Low);
	\draw[-{Latex[width=3mm,length=3mm]}] (root-NE) -- node [sloped,pos=0.8]{\LARGE{High}} (root-NE-High);
	 \draw[-{Latex[width=3mm,length=3mm]}] (root-SE) -- node [sloped,pos=0.8]{\LARGE{Low}}  (root-SE-Low);
	\draw[-{Latex[width=3mm,length=3mm]}] (root-SE) -- node [sloped,pos=0.8]{\LARGE{High}} (root-SE-High);
	\draw[-{Latex[width=3mm,length=3mm]}] (root-WE) -- node [sloped,pos=0.8]{\LARGE{Low}}  (root-WE-Low);
	\draw[-{Latex[width=3mm,length=3mm]}] (root-WE) -- node [sloped,pos=0.8]{\LARGE{High}} (root-WE-High);
	\draw[-{Latex[width=3mm,length=3mm]}] (root-EE-Low) -- node [sloped,pos=0.8]{\LARGE{Low}}  (root-EE-Low-Low);
	\draw[-{Latex[width=3mm,length=3mm]}] (root-EE-Low) -- node [sloped,pos=0.8]{\LARGE{High}} (root-EE-Low-High);
	\draw[-{Latex[width=3mm,length=3mm]}] (root-EE-High) -- node [sloped,pos=0.8]{\LARGE{Low}}  (root-EE-High-Low);
	\draw[-{Latex[width=3mm,length=3mm]}] (root-EE-High) -- node [sloped,pos=0.8]{\LARGE{High}} (root-EE-High-High);
	\draw[-{Latex[width=3mm,length=3mm]}]  (root-NE-Low) -- node [sloped,pos=0.8]{\LARGE{Low}}  (root-NE-Low-Low);
	\draw[-{Latex[width=3mm,length=3mm]}]  (root-NE-Low) -- node [sloped,pos=0.8]{\LARGE{High}} (root-NE-Low-High);
	\draw[-{Latex[width=3mm,length=3mm]}]  (root-NE-High) -- node [sloped,pos=0.8]{\LARGE{Low}}  (root-NE-High-Low);
	\draw[-{Latex[width=3mm,length=3mm]}] (root-NE-High) -- node [sloped,pos=0.8]{\LARGE{High}} (root-NE-High-High);
	\draw[-{Latex[width=3mm,length=3mm]}]  (root-SE-Low) -- node [sloped,pos=0.8]{\LARGE{Low}}  (root-SE-Low-Low);
	\draw[-{Latex[width=3mm,length=3mm]}]  (root-SE-Low) -- node [sloped,pos=0.8]{\LARGE{High}} (root-SE-Low-High);
	\draw[-{Latex[width=3mm,length=3mm]}]  (root-SE-High) -- node [sloped,pos=0.8]{\LARGE{Low}}  (root-SE-High-Low);
	\draw[-{Latex[width=3mm,length=3mm]}]  (root-SE-High) -- node [sloped,pos=0.8]{\LARGE{High}} (root-SE-High-High);
	\draw[-{Latex[width=3mm,length=3mm]}]  (root-WE-Low) -- node [sloped,pos=0.8]{\LARGE{Low}}  (root-WE-Low-Low);
	\draw[-{Latex[width=3mm,length=3mm]}]  (root-WE-Low) -- node [sloped,pos=0.8]{\LARGE{High}} (root-WE-Low-High);
	\draw[-{Latex[width=3mm,length=3mm]}] (root-WE-High) -- node [sloped,pos=0.8]{\LARGE{Low}}  (root-WE-High-Low);
	\draw[-{Latex[width=3mm,length=3mm]}]  (root-WE-High) -- node [sloped,pos=0.8]{\LARGE{High}} (root-WE-High-High);
	\draw[-{Latex[width=3mm,length=3mm]}]  (root-EE-Low-Low) -- node [sloped,pos=0.8]{\LARGE{Low}} (root-EE-Low-Low-High);
	\draw[-{Latex[width=3mm,length=3mm]}]  (root-EE-Low-Low) -- node [sloped,pos=0.8]{\LARGE{Medium}} (root-EE-Low-Low-Medium);
	\draw[-{Latex[width=3mm,length=3mm]}]  (root-EE-Low-Low) -- node [sloped,pos=0.8]{\LARGE{High}} (root-EE-Low-Low-Low);
	\draw[-{Latex[width=3mm,length=3mm]}]  (root-EE-Low-High) -- node [sloped,pos=0.8]{\LARGE{Low}} (root-EE-Low-High-High);
	\draw[-{Latex[width=3mm,length=3mm]}]  (root-EE-Low-High) -- node [sloped,pos=0.8]{\LARGE{Medium}} (root-EE-Low-High-Medium);
	\draw[-{Latex[width=3mm,length=3mm]}]  (root-EE-Low-High) -- node [sloped,pos=0.8]{\LARGE{High}} (root-EE-Low-High-Low);
	\draw[-{Latex[width=3mm,length=3mm]}]  (root-EE-High-Low) -- node [sloped,pos=0.8]{\LARGE{Low}}  (root-EE-High-Low-High);
	\draw[-{Latex[width=3mm,length=3mm]}]  (root-EE-High-Low) -- node [sloped,pos=0.8]{\LARGE{Medium}} (root-EE-High-Low-Medium);
	\draw[-{Latex[width=3mm,length=3mm]}]  (root-EE-High-Low) -- node  [sloped,pos=0.8]{\LARGE{High}} (root-EE-High-Low-Low);
	\draw[-{Latex[width=3mm,length=3mm]}]  (root-EE-High-High) -- node [sloped,pos=0.8]{\LARGE{Low}}  (root-EE-High-High-High);
	\draw[-{Latex[width=3mm,length=3mm]}]  (root-EE-High-High) -- node [sloped,pos=0.8]{\LARGE{Medium}} (root-EE-High-High-Medium);
	\draw[-{Latex[width=3mm,length=3mm]}]  (root-EE-High-High) -- node  [sloped,pos=0.8]{\LARGE{High}} (root-EE-High-High-Low);
	\draw[-{Latex[width=3mm,length=3mm]}]  (root-NE-Low-Low) -- node [sloped,pos=0.8]{\LARGE{Low}}  (root-NE-Low-Low-High);
	\draw[-{Latex[width=3mm,length=3mm]}]  (root-NE-Low-Low) -- node [sloped,pos=0.8]{\LARGE{Medium}} (root-NE-Low-Low-Medium);
	\draw[-{Latex[width=3mm,length=3mm]}]  (root-NE-Low-Low) -- node  [sloped,pos=0.8]{\LARGE{High}} (root-NE-Low-Low-Low);
	\draw[-{Latex[width=3mm,length=3mm]}]  (root-NE-Low-High) -- node [sloped,pos=0.8]{\LARGE{Low}}  (root-NE-Low-High-High);
	\draw[-{Latex[width=3mm,length=3mm]}]  (root-NE-Low-High) -- node [sloped,pos=0.8]{\LARGE{Medium}} (root-NE-Low-High-Medium);
	\draw[-{Latex[width=3mm,length=3mm]}]  (root-NE-Low-High) -- node  [sloped,pos=0.8]{\LARGE{High}} (root-NE-Low-High-Low);
	\draw[-{Latex[width=3mm,length=3mm]}]  (root-NE-High-Low) -- node [sloped,pos=0.8]{\LARGE{Low}}  (root-NE-High-Low-High);
	\draw[-{Latex[width=3mm,length=3mm]}]  (root-NE-High-Low) -- node [sloped,pos=0.8]{\LARGE{Medium}} (root-NE-High-Low-Medium);
	\draw[-{Latex[width=3mm,length=3mm]}]  (root-NE-High-Low) -- node  [sloped,pos=0.8]{\LARGE{High}} (root-NE-High-Low-Low);
	\draw[-{Latex[width=3mm,length=3mm]}]  (root-NE-High-High) -- node [sloped,pos=0.8]{\LARGE{Low}}  (root-NE-High-High-High);
	\draw[-{Latex[width=3mm,length=3mm]}]  (root-NE-High-High) -- node [sloped,pos=0.8]{\LARGE{Medium}} (root-NE-High-High-Medium);
	\draw[-{Latex[width=3mm,length=3mm]}]  (root-NE-High-High) -- node  [sloped,pos=0.8]{\LARGE{High}} (root-NE-High-High-Low);
	\draw[-{Latex[width=3mm,length=3mm]}]  (root-SE-Low-Low) -- node [sloped,pos=0.8]{\LARGE{Low}}  (root-SE-Low-Low-High);
	\draw[-{Latex[width=3mm,length=3mm]}]  (root-SE-Low-Low) -- node [sloped,pos=0.8]{\LARGE{Medium}} (root-SE-Low-Low-Medium);
	\draw[-{Latex[width=3mm,length=3mm]}]  (root-SE-Low-Low) -- node  [sloped,pos=0.8]{\LARGE{High}} (root-SE-Low-Low-Low);
	\draw[-{Latex[width=3mm,length=3mm]}]  (root-SE-Low-High) -- node [sloped,pos=0.8]{\LARGE{Low}}  (root-SE-Low-High-High);
	\draw[-{Latex[width=3mm,length=3mm]}]  (root-SE-Low-High) -- node [sloped,pos=0.8]{\LARGE{Medium}} (root-SE-Low-High-Medium);
	\draw[-{Latex[width=3mm,length=3mm]}]  (root-SE-Low-High) -- node  [sloped,pos=0.8]{\LARGE{High}} (root-SE-Low-High-Low);
	\draw[-{Latex[width=3mm,length=3mm]}]  (root-SE-High-Low) -- node [sloped,pos=0.8]{\LARGE{Low}}  (root-SE-High-Low-High);
	\draw[-{Latex[width=3mm,length=3mm]}]  (root-SE-High-Low) -- node [sloped,pos=0.8]{\LARGE{Medium}} (root-SE-High-Low-Medium);
	\draw[-{Latex[width=3mm,length=3mm]}]  (root-SE-High-Low) -- node  [sloped,pos=0.8]{\LARGE{High}} (root-SE-High-Low-Low);
	\draw[-{Latex[width=3mm,length=3mm]}]  (root-SE-High-High) -- node [sloped,pos=0.8]{\LARGE{Low}}  (root-SE-High-High-High);
	\draw[-{Latex[width=3mm,length=3mm]}]  (root-SE-High-High) -- node [sloped,pos=0.8]{\LARGE{Medium}} (root-SE-High-High-Medium);
\draw[-{Latex[width=3mm,length=3mm]}]  (root-SE-High-High) -- node  [sloped,pos=0.8]{\LARGE{High}} (root-SE-High-High-Low);
\draw[-{Latex[width=3mm,length=3mm]}]  (root-WE-Low-Low) -- node [sloped,pos=0.8]{\LARGE{Low}}  (root-WE-Low-Low-High);
	\draw[-{Latex[width=3mm,length=3mm]}] (root-WE-Low-Low) -- node [sloped,pos=0.8]{\LARGE{Medium}} (root-WE-Low-Low-Medium);
\draw[-{Latex[width=3mm,length=3mm]}] (root-WE-Low-Low) -- node  [sloped,pos=0.8]{\LARGE{High}} (root-WE-Low-Low-Low);
\draw[-{Latex[width=3mm,length=3mm]}]  (root-WE-Low-High) -- node [sloped,pos=0.8]{\LARGE{Low}}  (root-WE-Low-High-High);
	\draw[-{Latex[width=3mm,length=3mm]}]  (root-WE-Low-High) -- node [sloped,pos=0.8]{\LARGE{Medium}} (root-WE-Low-High-Medium);
	\draw[-{Latex[width=3mm,length=3mm]}]  (root-WE-Low-High) -- node  [sloped,pos=0.8]{\LARGE{High}} (root-WE-Low-High-Low);
	\draw[-{Latex[width=3mm,length=3mm]}]  (root-WE-High-Low) -- node [sloped,pos=0.8]{\LARGE{Low}}  (root-WE-High-Low-High);
	\draw[-{Latex[width=3mm,length=3mm]}] (root-WE-High-Low) -- node [sloped,pos=0.8]{\LARGE{Medium}} (root-WE-High-Low-Medium);
	\draw[-{Latex[width=3mm,length=3mm]}]  (root-WE-High-Low) -- node  [sloped,pos=0.8]{\LARGE{High}} (root-WE-High-Low-Low);
\draw[-{Latex[width=3mm,length=3mm]}] (root-WE-High-High) -- node [sloped,pos=0.8]{\LARGE{Low}}  (root-WE-High-High-High);
	\draw[-{Latex[width=3mm,length=3mm]}] (root-WE-High-High) -- node [sloped,pos=0.8]{\LARGE{Medium}} (root-WE-High-High-Medium);
	\draw[-{Latex[width=3mm,length=3mm]}] (root-WE-High-High) -- node [sloped,pos=0.8]{\LARGE{High}} (root-WE-High-High-Low);
\end{tikzpicture}
}
    \caption{Staged tree for the railway service satisfaction example whose staging represents all probabilities equalities in Figure \ref{fig:bn1} and Table \ref{table:1}.}
    \label{fig:staged}
\end{figure}
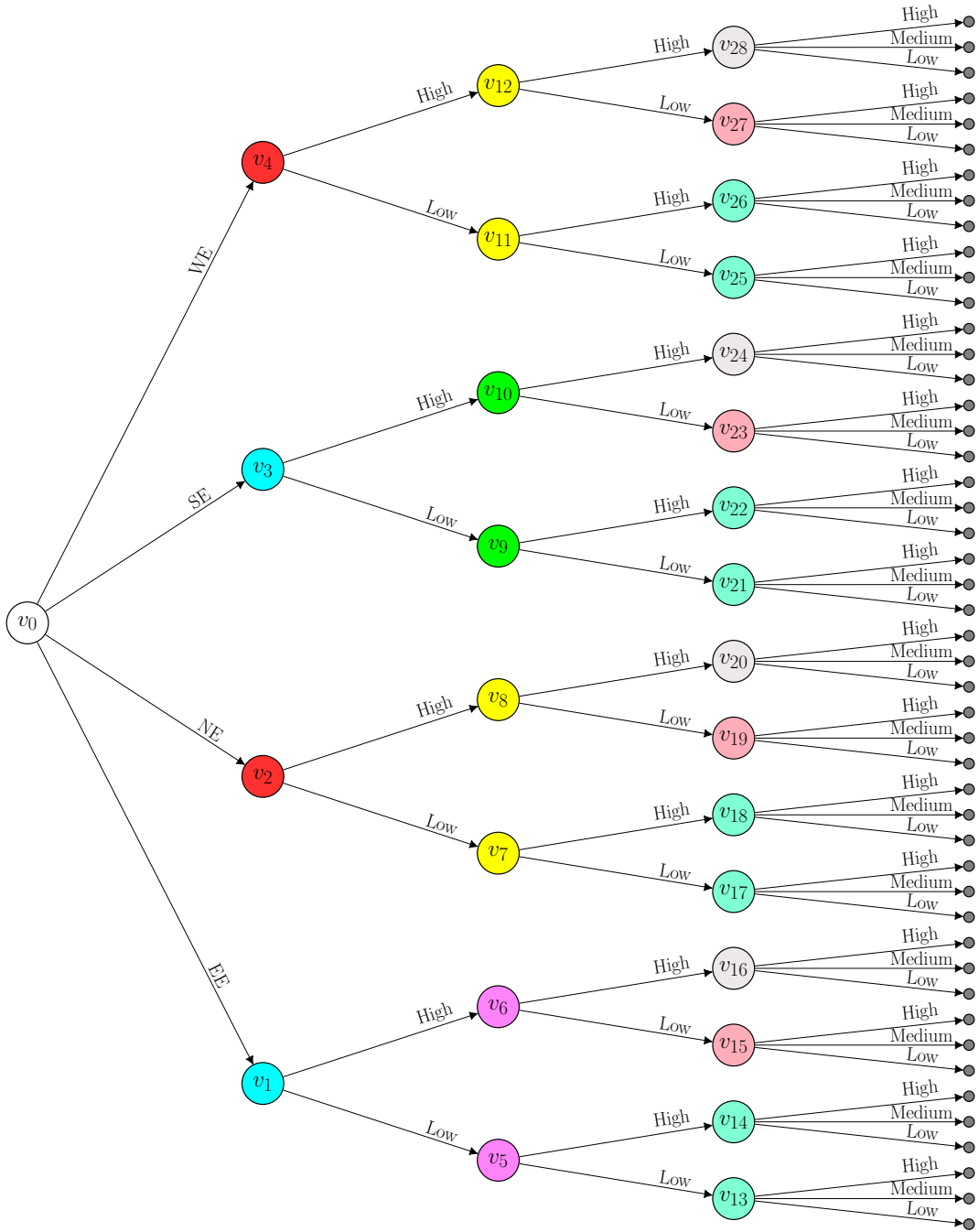

As an illustration, consider the tree in Figure \ref{fig:staged} (the coloring of the vertices is, for now, irrelevant) representing the sample space of our running example in railway service satisfaction. Recall that the variables of interest are Country, Length, Income, and Satisfaction; and suppose we fix this ordering of the variables. The root of the tree $v_0$ is associated with the first variable Country, and the edges emanating from $v_0$ represent the possible values Country can take, namely EE, NE, SE, WE. The probability associated with these four edges (not reported here) must sum up to one. Now consider the vertex $v_1$. The edges emanating from $v_1$ denote the conditional probability of Length given Country=EE, and again, their probabilities must add up to one. All non-leaf vertices at the same depth in the tree are associated with conditional distributions from the same variable. Because of the standard chain rule of probabilities, the product of the edge probabilities in a root-to-leaf path is the probability of an atomic event. Consider the lowest path: this represents the event Country=EE, Length=Low, Income=Low, Satisfaction=Low.

Conditional independences are visualized in the tree by a coloring of the non-leaf vertices. Two vertices that are assigned the same color are such that their edge probabilities are the same. Furthermore, two equally-colored vertices are said to be in the same \textit{stage}. Considering again Figure \ref{fig:staged}, it can be seen that $v_1$ and $v_3$ are given the same color. This means that $P(\textnormal{Length=High}|\textnormal{Country=EE})=P(\textnormal{Length=High}|\textnormal{Country=SE})$ (and similarly for Length=Low): a marginal partial independence. Considering vertices at depth two, it can be noticed that the stages are $\{v_5,v_6\},\{v_7,v_8\},\{v_9,v_{10}\}$, and $\{v_{11},v_{12}\}$. This implies, for instance (vertices  $\{v_5,v_6\}$), that the conditional distribution of Income given Country=EE and Length=Low is the same as the one given Country=EE and Length=High. However, it can be seen that this happens for every value of Country (EE, NE, SE, WE). Altogether, this means that Income and Length are conditionally independent given Country: a symmetric, traditional conditional independence.  By investigating further the coloring of the staged tree in Figure \ref{fig:staged} additional symmetric and non-symmetric independences can be identified.

This example clearly illustrates that staged trees can embed both symmetric and non-symmetric types of independence via their coloring. BNs can be seen as a subclass of staged trees \citep{smith2008conditional,varando2021staged} since all BNs can be represented as staged trees, but not vice-versa since staged trees can graphically represent non-symmetric independences. By a thorough investigation of the staged tree in Figure \ref{fig:staged}, it can be noticed that its staging is such that the only equalities between probabilities it enforces are the two symmetric conditional independences from the DAG in Figure \ref{fig:bn1} and the constraints in the CPTs in Table \ref{table:1}. Therefore, the staged tree in Figure \ref{fig:staged} gives a complete graphical representation of the BN, including the equalities enforced in its CPTs. 

Although staged trees could be expert-elicited, our focus here is on data-driven structural learning algorithms. In the context of staged trees, these require three steps: (i) learning an optimal ordering of the variables; (ii) learning the staging of the non-leaf vertices; (iii) learning the edge probabilities. Step (iii) is similarly conducted as in BNs using either frequentist or Bayesian approaches. Step (i) is computationally challenging due to the factorial explosion of the number of orderings. For a small number of variables, the ordering can be found using a dynamic programming approach \citep{cowell2014causal,leonelli2023context}. In more complex scenarios, a fraction of the space of orderings could be explored \citep{leonelli2023learning}. Step (ii) is discussed in detail in Section \ref{sec:structural}.

One important consideration is that structural learning of generic staged trees is hard, due to the explosion of the model search space as the number of variables increases \citep[see e.g.][]{duarte2021representation}. For this reason, recent research has focused on sub-classes of staged tree models: \citet{carli2020new} defined naive staged trees that have the same number of parameters of a naive BN over the same variables; \citet{leonelli2022structural} considered simple staged trees which have a constrained type of partitioning of the vertices;  \citet{duarte2021representation} defined CStrees which only embed symmetric and context-specific types of independence; \citet{leonelli2022highly} introduced \textit{k-parents staged trees}, which limit the number of variables that can have a direct influence on another.

A wide array of methods to efficiently investigate real-world applications are now available for staged trees, including user-friendly software \citep{Carli2022,walley2023cegpy}, inferential and sensitivity routines \citep{gorgen2015differential,leonelli2019sensitivity,thwaites2008propagation}, dealing with missing data \citep{barclay2014chain}, causal reasoning \citep{thwaites2010causal} and identification of equivalence classes \citep{gorgen2018discovery}, to name a few. Such techniques are, in general, not available for other graphical models embedding non-symmetric independences, thus making staged trees a viable as well as efficient option for applied analyses.

\subsection{Asymmetry-labeled DAGs}

An additional challenge in modeling categorical data with staged trees is that the sample space, and therefore the size of the tree to be plotted, grows super-exponentially with the number of variables. In our simple example with one quaternary, one ternary, and two binary variables, the tree becomes already relatively big. In practice, it becomes almost impossible to visualize it and, therefore, graphically assess the learned dependence structure with more than seven binary variables.

To address this visualization challenge, \citet{varando2021staged} introduced a compression of a staged tree into a DAG, called \textit{asymmetry-labeled DAG (ALDAG)}  having the following properties: (i) $X_i\indep X_j |X_k$ is implied by the staging of the tree if and if only $X_i$ and $X_j$ are d-separated by $X_k$ in the ALDAG; (ii) the ALDAG is minimal, in the sense that there are no other DAGs with a smaller number of edges respecting property (i); (iii) the edges of the ALDAG are labeled/colored to denote the type of non-symmetric independence existing between the relevant variables. 

\begin{figure}
\centering
\scalebox{0.8}{
\begin{tikzpicture}[auto, scale=10,
	Country_NA/.style={circle,inner sep=0.5mm,minimum size=2.5cm,draw,black,very thick,text=black}]
\node [Country_NA] (1) at (0,0.25){Country};
\node [Country_NA] (2) at (0.5,0){Length};
\node [Country_NA](3) at (0.5,0.5){Income};
\node [Country_NA] (4) at (1,0.25){Satisfaction};
\draw[-{Latex[width=3mm,length=3mm]}, line width = 1.1pt,green] (1) -- (2);
\draw[-{Latex[width=3mm,length=3mm]}, line width = 1.1pt,green] (1) -- (3);
\draw[-{Latex[width=3mm,length=3mm]}, line width = 1.1pt,red] (3) -- (4);
\draw[-{Latex[width=3mm,length=3mm]}, line width = 1.1pt] (2) -- (4);
\end{tikzpicture}
}
\caption{ALDAG associated with the staged tree in Figure \ref{fig:staged}, or, equivalently, with the DAG in Figure \ref{fig:bn1} with CPTs in Table \ref{table:1}. Labels: black - symmetric dependence; red - context-specific; green - partial.  \label{fig:aldag}}
\end{figure}
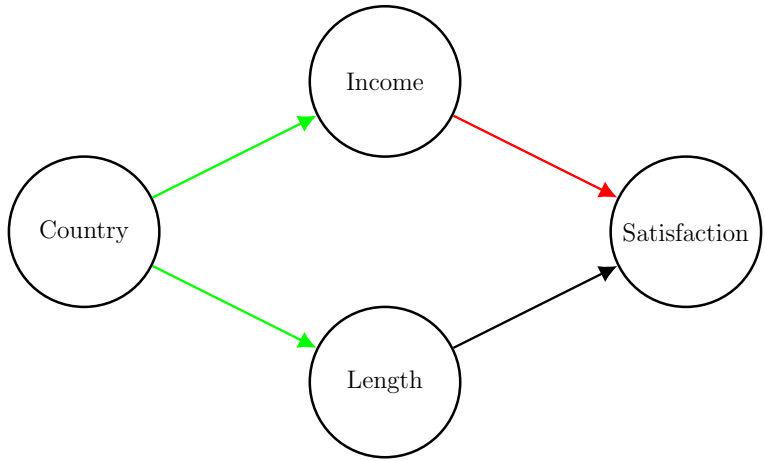

Another interpretation of ALDAGs is as an embellishment of a standard BN, letting edges denote the existence of additional probability equalities in its CPTs. To illustrate this, consider the running railway service satisfaction example with BN in Figure \ref{fig:bn1} and CPTs in Table \ref{table:1}. As already noticed, the CPTs include additional probability constraints that cannot be visualized from the DAG. The ALDAG representing this extra information is reported in Figure \ref{fig:aldag}, which, by construction, has the same edge set as the original DAG. The CPT of Length includes the partial independences $P(\textnormal{Length=High}|\textnormal{Country=EE})=P(\textnormal{Length=High}|\textnormal{Country=SE})$ and $P(\textnormal{Length=High}|\textnormal{Country=WE})=P(\textnormal{Length=High}|\textnormal{Country=NE})$: thus the edge is labeled partial. A similar observation can be made for the CPT of Income. Concerning the CPT of Satisfaction, we already noticed that the probability distribution of Satisfaction does not depend on Income when Length=Low. For this reason, the edge from Income to Satisfaction is given the label context-specific. On the other hand, if the value of Income is kept fixed to either High or Low,  changing Length from Low to High has an effect on the probability of Satisfaction. For this reason, the edge from Length to Satisfaction is given a symmetric label since this is the type of relationship that graphically can be described by the DAG and symmetric conditional independence.

The ALDAG in Figure \ref{fig:aldag} would also be obtained by applying the compression algorithm of \citet{varando2021staged} to the staged tree in Figure \ref{fig:staged}, since it specifically embeds the probability equalities defined by the DAG and CPTs considered before. The conversion of a staged tree into its associated ALDAG has the advantage that the many routines available for inference over BNs, for instance, the already-mentioned fast inferential algorithms and what-if analyses, can be straightforwardly used for staged trees as well. In our applications below, we showcase the use of what-if analyses over the associated ALDAG of a learned staged tree.

\citet{leonelli2023learning} noticed that for staged trees learned from observational data, their associated ALDAGs are usually complete unless some sparsity is imposed during the structural learning algorithm. The interpretational advantages of compressing the tree would basically be lost with a complete ALDAG. For this reason, \citet{leonelli2022highly} introduced $k$-parents staged that are staged trees such that the maximum in-degree in the associated ALDAG is $k$. Restricting the number of parents makes sense from an applied point of view since, most often, only a limited number of variables can be expected to have a direct influence on another. The option of setting a maximum number of parents is also available in the standard \texttt{bnlearn} software \citep{Scutari2010}. Furthermore, by considering $k$-parents staged treed for a small $k$, the model search space is drastically reduced, and computationally efficient structural learning algorithms are available.

Limiting the number of parents has the advantage that, by applying the ordered Markov condition over the ALDAG, irrelevant variables can be removed from the visualization of the staging of a variable of interest. \citet{varando2021staged} introduced the so-called \textit{dependence subtree}, which, for each variable, shows its staging using only the relevant parent variables. By visualizing the ALDAG and the dependence subtrees, the whole, complex, non-symmetric dependence structure can be investigated even in situations with a large number of variables.  

\section{Robust learning algorithms for staged trees}
\label{sec3}

It is now common practice in applied structural learning of BNs to assess the strength of the learned relationships, i.e. \textit{validate} the learned network. The most common approach is to resample with replacement the data via non-parametric bootstrap and to estimate a BN for each of the resampled datasets \citep{caravagna2021learning,friedman1999data}. Edges that appear a frequency of times above some threshold are then retained in the final DAG \citep{denis}. Although there are now theoretical methods to choose such a threshold \citep{scutari}, most often it is manually fixed at some pre-specified level \citep[e.g.][]{briganti2022using}. Furthermore, the validity of the model is often assessed by evaluating out-of-sample predictions using a $k$-fold cross-validation \citep{liew2022short}. The above-mentioned resampling strategy is then applied within each iteration of the cross-validation. Both resampling and cross-validation can be easily implemented in the learning of BNs using the \texttt{bn.boot} and \texttt{bn.cv} functions of the \texttt{bnlearn} R package.

Despite their widespread use for learning BNs, such techniques have not been applied for learning staged trees from data. The quality of a staged tree model is most often evaluated over the training dataset without an out-of-sample assessment of the validity of the model. Before discussing how this could be implemented in practice, we review standard structural learning algorithms for staged trees.

\subsection{Structural learning for staged trees}
\label{sec:structural}

In Section \ref{sec:staged} above we started discussing the learning of staged trees from data. The first step is choosing an ordering of the variables in the tree. For a small number of variables, all possible orderings can be evaluated by taking advantage of a dynamic programming approach (implemented in the \texttt{search\_best} function in the \texttt{stagedtrees} R package) \citep{cowell2014causal,leonelli2023context}. Otherwise, the space of possible orderings can be initially pruned, for instance by only selecting orders compatible with a learned BN, so that the optimal ordering selection can also be performed when more variables are studied \citep{leonelli2023learning}. However, most often, the order is selected using common knowledge, expert opinion, and, if available, the natural causal ordering of the variables.

Once an ordering has been selected, an optimal staging of the vertices has to be learned. By construction, only vertices at the same depth of the tree, i.e. vertices representing the conditional distributions of the same variable, can be merged in the same stage. Practically, this comes down to finding an optimal clustering of the vertices by exploring the space of vertices' partitions. Although distance-based \citep{carli2020new} and $k$-means \citep{silander2013dynamic} methods have been proposed, staged trees are most commonly learned using greedy hill-climbing techniques which optimize a model score \citep[most often the BIC as discussed in][]{gorgen2022curved}. In its most general form (implemented in the \texttt{stages\_hc} function of the \texttt{stagedtrees} R package) every possible merging and splitting of stages is considered at each iteration. 

The most common greedy search routine is, on the other hand, the \textit{backward hill climbing} (or agglomerative hierarchical clustering) algorithm \citep{freeman2011bayesian} which, starting from the tree where every vertex is in its own stage, at each iteration merges the pair of stages leading to the best score increase until no improvement is found. This routine is implemented in the \texttt{stages\_bhc} function of the \texttt{stagedtrees} R package and is henceforth referred to as BHC. Notice that the learning of the staging can be performed independently and in parallel for vertices at different depths in the tree because the likelihood (and hence most model's scores as BIC) separates across the variables.

Because of the size of the data applications we study in Section \ref{sec4}, we will also use the techniques to learn $k$-parents staged trees introduced in \citet{leonelli2023learning} and reported in the appendix. For each variable, these first select $k$ parents that maximize the conditional mutual information and then run the BHC to select an optimal staging. Notice that the resulting tree is guaranteed to be $k$-parent since no additional parents can be added during the learning of the staging.

\subsection{Robust learning of a variables' ordering}
\label{sec:order}
We now describe how non-parametric bootstrap resampling techniques can be used in the context of learning staged trees from data, starting by deciding on a variables' ordering. Suppose a dataset $\mathcal{D}$ of $N$ observations of the variables of interest $X_1,\dots, X_p$ is available. We construct $M$ synthetic versions of $\mathcal{D}$ each of size $N$, called $\mathcal{D}^{(1)},\dots,\mathcal{D}^{(M)}$, using non-parametric bootstrap.

Suppose we chose a specific learning algorithm for the staging of the tree (e.g. BHC). For $i=1,\dots, M$, we apply the dynamic programming approach using $\mathcal{D}^{(i)}$ to learn an optimal ordering of the variables $X_\sigma^{(i)}$. This gives us $M$ variables' orderings.

For every pair $(j,k)$, $j,k=1,\dots,p$, we compute the frequency that $X_j$ preceded $X_k$ in the order, i.e.
\[
N_{jk} =\frac{1}{M}\sum_{i=1}^M \mathds{1}_{X_{\sigma(j)}^{(i)}\prec X_{\sigma(k)}^{(i)}} 
\]
The final ordering of the variables $X_{\sigma}$ is then uniquely defined by the $N_{jk}\geq 0.5$.\footnote{In the rare event of ties when $N_{jk}=0.5$ the ordering of the associated variables can be randomly chosen.}

Of course, this routine has a factorial complexity and can be implemented only for a small number of variables. However, we showcase in Section \ref{sec:rail} that the order selection can also be performed in complex scenarios by using expert information to group the variables.

\subsection{Robust learning with a fixed ordering}

A non-parametric bootstrap is then used to estimate the staging of the tree. Again, assume that a staging learning algorithm has been chosen, $\mathcal{D}^{(1)},\dots,\mathcal{D}^{(M)}$ synthetic copies of $\mathcal{D}$ have been generated, and that an optimal ordering $X_{\sigma}$ has been fixed. For each $\mathcal{D}^{(i)}$, we estimate and optimal staged tree $T^{(i)}$ with an optimal variable ordering $X_{\sigma}$. Call $U^{(i)}_{\sigma(j)}$ the staging of the vertices at depth $j$ in the i-th bootstrap replicate, $j=1,\dots,p-1$, $i=1,\dots,M$.

Given the $M$ learned stagings $U^{(i)}_{\sigma(j)}$, $j=1,\dots,M$, a method to combine them into a unique one must be adopted. This comes down to finding an optimal way to summarize $M$ different partitions of the same set, a problem that has been addressed in Bayesian clustering \citep[e.g.][]{wade2023bayesian}. In this work, we apply a novel methodology that resembles the one to choose a final BN after bootstrapping:
\begin{itemize}
    \item Create a matrix $Z_j$ whose columns are $U^{(1)}_{\sigma(j)},\dots,U^{(1)}_{\sigma(j)}$;
    \item Compute the pairwise dissimilarity matrix $D_j$, giving for each pair of elements the frequency of times they are not in the same subset \citep[the \texttt{psm} function from the \texttt{salso} R package is used][]{dahl2022search};
    \item A standard agglomerative hierarchical clustering is run over $D_j$ using the \texttt{hclust} R function;
    \item The associated dendrogram is cut at a pre-specified height, giving the final staging $U_{\sigma(j)}^*$.
\end{itemize}

The averaged staged tree model is then $T^*$ with staging $U_{\sigma(j)}^*$, $j=1,\dots,p-1.$ Notice that if the above method is used in conjunction with algorithms to learn $k$-parents staged trees, the resulting tree $T^*$ is not necessarily within the same model class, and its associated ALDAG can have a maximum in-degree larger than $k$. However, as illustrated by our applications below, the resulting staged tree and ALDAG are still sparse.

As our routine computes $M$ different staged trees, they can each be converted into their ALDAG representation $G^{(i)}$. As in BNs, these can be used to assess the strength of the relationship between every pair of variables. Furthermore, as they are ALDAGs, the strength of symmetric and non-symmetric relationships can be assessed by computing the frequency of each possible type of edge.

\section{Applications}
\label{sec4}

We now showcase the use of our newly defined methods to analyze customer satisfaction surveys. BNs have been used extensively for this task \citep[e.g.][]{cugnata2014bayesian,salini2009bayesian}. Here, we demonstrate the additional insights staged trees and ALDAGs can provide to untangle complex dependence patterns between the factors that affect customer satisfaction.

\subsection{Airline passengers' satisfaction}
\label{sec:air}
We first analyze a simpler dataset considered in \citet{cugnata2016bayesian} about the satisfaction of airlines' passengers. The questionnaire contains questions on the passengers’ satisfaction with their overall experience and six specific service dimensions (departure, booking, check-in, cabin environment, cabin crew, and meal). The evaluation of each item is based on a four-point scale (from extremely dissatisfied to extremely satisfied), which, for simplicity, is merged into two levels (low/high). A total of 9720 responses are available. The aim of the analysis is to evaluate the importance of these six service dimensions on the overall experience. PGMs provide an intuitive platform for this type of analysis since these service dimensions cannot be assumed independent.

We consider five different structural learning algorithms for staged trees: the BHC and the $k$-parent staged trees learning algorithm for $k=1,2,3,4$. We start by choosing the best ordering for each algorithm among the six specific service dimensions, while the overall experience is fixed as the last variable in the ordering since the aim is to understand the effect the dimensions have on it. Table \ref{table:order_air} reports the learned order for each of the used algorithms using the procedure outlined in Section \ref{sec:order}. It can be seen that the first three variables are equally ordered for all methods, while the last three may have a different ordering depending on the algorithm used.

\begin{table}
\begin{center}
\begin{tabular}{c|ccccc}
\toprule
Algorithm & BHC & 1-parent & 2-parents & 3-parents & 4-parents \\
\midrule
1st & crew & crew & crew & crew & crew \\
2nd & cabin & cabin & cabin & cabin & cabin \\
3rd & meal & meal & meal & meal & meal \\
4th & booking & departure & departure & departure & departure \\
5th & departure & check-in & booking & check-in & booking \\
6th & check-in & booking & check-in & booking & check-in\\ 
\bottomrule
\end{tabular}
\end{center}
\caption{Ordering of the variables for different staged tree learning algorithms using the bootstrap approach over the airline dataset. \label{table:order_air}}
\end{table}

\begin{figure}
\begin{center}
\includegraphics[scale=0.6]{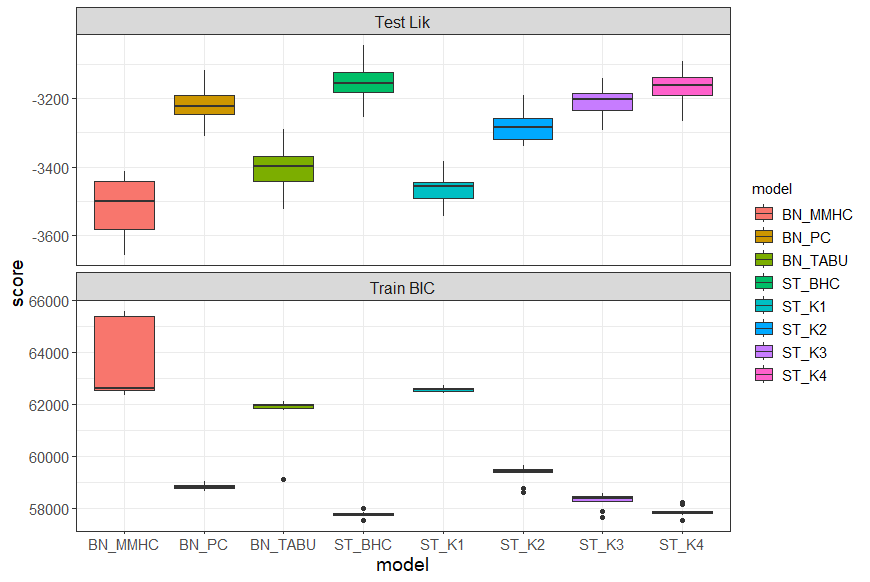}
\end{center}
\caption{Boxplots of BIC scores (train dataset) and log-likelihood (test dataset) from a 10-fold cross-validation over the airline dataset.}
\label{fig:cv1}
\end{figure}

Having selected the orderings, we then run a 10-fold cross-validation for the five staged tree algorithms of above and three BN structural learning algorithms for BNs: MMHC, PC and tabu implemented in the \texttt{bnlearn} R package. For each of the 10 folds, we run a non-parametric bootstrap of 200 iterations and the thresholds for both BNs and staged trees were set at 0.5. For each fold, we then computed the BIC of the models over the training data and the predictive log-likelihood over the test data. Figure \ref{fig:cv1} reports the results. Staged trees learned with the BHC (ST\_BHC) and the algorithm for 4-parents staged trees (ST\_K4) outperform all BN approaches in both fitting and predictive tasks. Of course, since it is considered the most general class of models, ST\_BHC outperforms ST\_K4, but the difference appears to be marginal. Thus, the presence of asymmetric dependence is critical to fully understanding the relationship between the service dimensions and the overall experience.

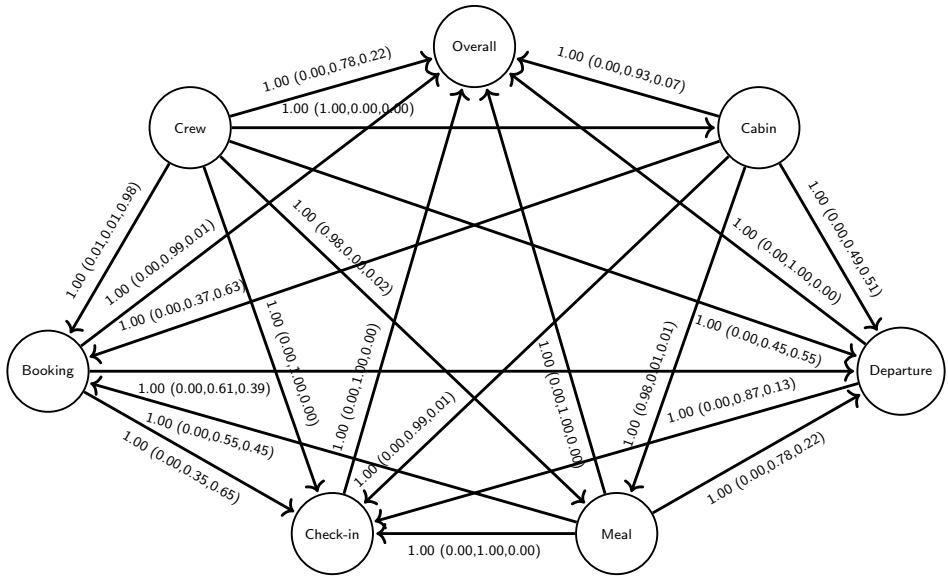
\begin{figure}
\begin{center}
\scalebox{0.9}{
  \begin{tikzpicture}[
        shorten >=1pt, auto, thick,
        node distance=3cm,
    main node/.style={circle,draw,font=\sffamily\tiny, minimum size = 1.2cm}
                            ]
      \node[main node] (crew) at (1*\xx,2.5*\yy) {Crew};
      \node[main node] (cabin) at (5*\xx,2.5*\yy) {Cabin};
	\node[main node] (departure) at (6*\xx,1*\yy) {Departure};
\node[main node] (meal) at (4*\xx,0*\yy) {Meal};
\node[main node] (checkin) at (2*\xx,0*\yy) {Check-in};
\node[main node] (booking) at (0*\xx,1*\yy){Booking};
\node[main node] (overall) at (3*\xx,3*\yy){Overall};
\draw[->,line width=1*\zz pt] 
  (crew)  -- node [sloped,swap,pos=0.85]{\sffamily\tiny{1.00 (0.00,0.45,0.55)}} (departure); 
  \draw[->,line width=1*\zz pt] 
  (crew)  -- node [sloped,pos=0.24]{\sffamily\tiny{1.00 (1.00,0.00,0.00)}} (cabin); 
    \draw[->,line width=1*\zz pt] 
  (crew)  -- node [sloped,above]{\sffamily\tiny{1.00 (0.01,0.01,0.98)}} (booking); 
      \draw[->,line width=1*\zz pt] 
  (crew)  -- node [sloped,pos=0.62]{\sffamily\tiny{1.00 (0.00,1.00,0.00)}} (checkin); 
        \draw[->,line width=1*\zz pt] 
  (crew)  -- node [sloped,above,pos=0.3]{\sffamily\tiny{1.00 (0.98,0.00,0.02)}} (meal); 
          \draw[->,line width=1*\zz pt] 
  (crew)  -- node [sloped]{\sffamily\tiny{1.00 (0.00,0.78,0.22)}} (overall); 
            \draw[->,line width=1*\zz pt] 
  (cabin)  -- node [sloped,pos=0.68]{\sffamily\tiny{1.00 (0.98,0.01,0.01)}} (meal); 
              \draw[->,line width=1*\zz pt] 
  (cabin)  -- node [sloped,pos=0.84]{\sffamily\tiny{1.00 (0.00,0.37,0.63)}} (booking); 
                \draw[->,line width=1*\zz pt] 
  (cabin)  -- node [sloped]{\sffamily\tiny{1.00 (0.00,0.49,0.51)}} (departure); 
                  \draw[->,line width=1*\zz pt] 
  (cabin)  -- node [sloped,pos=0.85]{\sffamily\tiny{1.00 (0.00,0.99,0.01)}} (checkin);
                  \draw[->,line width=1*\zz pt] 
  (cabin)  -- node [sloped]{\sffamily\tiny{1.00 (0.00,0.93,0.07)}} (overall);  
                    \draw[->,line width=1*\zz pt] 
  (meal)  -- node [sloped,pos=.74,below]{\sffamily\tiny{1.00 (0.00,0.55,0.45)}} (booking);  
                      \draw[->,line width=1*\zz pt] 
  (meal)  -- node [sloped,below]{\sffamily\tiny{1.00 (0.00,0.78,0.22)}} (departure);  
                        \draw[->,line width=1*\zz pt] 
  (meal)  -- node [sloped,below]{\sffamily\tiny{1.00 (0.00,1.00,0.00)}} (checkin);  
                        \draw[->,line width=1*\zz pt] 
  (meal)  -- node [sloped,below,pos=0.23]{\sffamily\tiny{1.00 (0.00,1.00,0.00)}} (overall);  
                          \draw[->,line width=1*\zz pt] 
  (booking)  -- node [sloped,below,pos=0.15]{\sffamily\tiny{1.00 (0.00,0.61,0.39)}} (departure);  
                            \draw[->,line width=1*\zz pt] 
  (booking)  -- node [sloped,below]{\sffamily\tiny{1.00 (0.00,0.35,0.65)}} (checkin);  
                            \draw[->,line width=1*\zz pt] 
  (booking)  -- node [sloped,pos=0.25]{\sffamily\tiny{1.00 (0.00,0.99,0.01)}} (overall);  
                              \draw[->,line width=1*\zz pt] 
  (departure)  -- node [sloped,above,pos=0.25]{\sffamily\tiny{1.00 (0.00,0.87,0.13)}} (checkin);  
                            \draw[->,line width=1*\zz pt] 
  (departure)  -- node [sloped,pos=0.25]{\sffamily\tiny{1.00 (0.00,1.00,0.00)}} (overall);  
                              \draw[->,line width=1*\zz pt] 
  (checkin)  -- node [sloped,pos=0.25]{\sffamily\tiny{1.00 (0.00,1.00,0.00)}} (overall);  
\end{tikzpicture}
}
\end{center}
\caption{Edge strengths of the ALDAGs learned with the ST\_BHC algorithm over the airline dataset. In parenthesis, the proportion of times edges are given labels symmetric, context-specific, and local.}
\label{fig:aldag_BHC}
\end{figure}

As common in BN applied structural learning \citep[e.g.][]{liew2022short}, we then applied the non-parametric bootstrap approach over the full dataset for the ST\_BHC and ST\_K4 models. Figure \ref{fig:aldag_BHC} reports the edge strengths for the ST\_BHC algorithm. As already noticed, unless sparsity is imposed in the learning of staged trees, the learned ALDAGs are fully connected, as evidenced by all weights being equal to one. In parenthesis, the proportion of times edges are of symmetric, context-specific, or local types is reported (since all variables are binary no partial dependence can exist). The data clearly shows the presence of non-symmetric dependence that BNs cannot model because of their assumption of symmetric conditional independence.

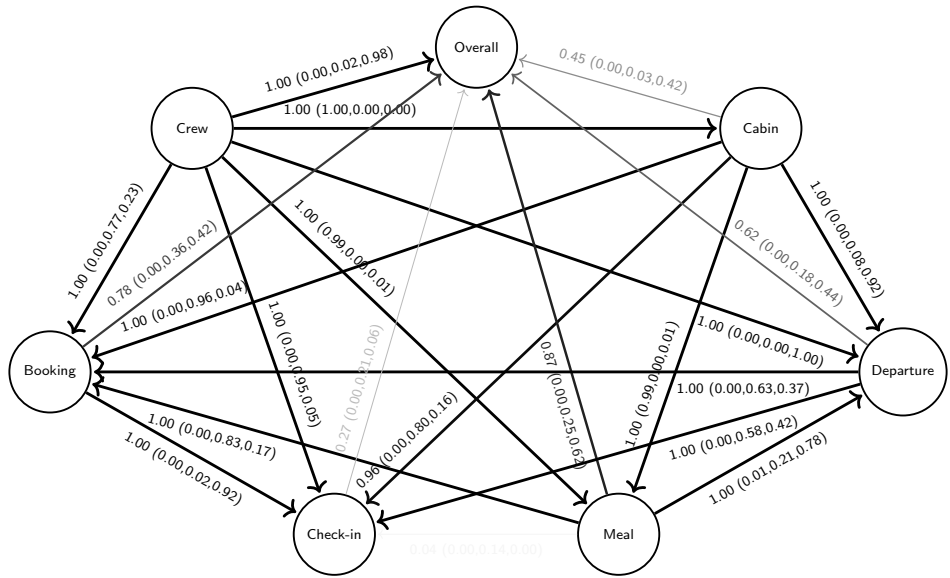
\begin{figure}
\begin{center}
\scalebox{0.9}{
  \begin{tikzpicture}[
        shorten >=1pt, auto, thick,
        node distance=3cm,
    main node/.style={circle,draw,font=\sffamily\tiny, minimum size = 1.2cm}
                            ]
      \node[main node] (crew) at (1*\xx,2.5*\yy) {Crew};
      \node[main node] (cabin) at (5*\xx,2.5*\yy) {Cabin};
	\node[main node] (departure) at (6*\xx,1*\yy) {Departure};
\node[main node] (meal) at (4*\xx,0*\yy) {Meal};
\node[main node] (checkin) at (2*\xx,0*\yy) {Check-in};
\node[main node] (booking) at (0*\xx,1*\yy){Booking};
\node[main node] (overall) at (3*\xx,3*\yy){Overall};
\draw[->,line width=1*\zz pt]  (crew)  -- node [sloped,swap,pos=0.85]{\sffamily\tiny{1.00 (0.00,0.00,1.00)}} (departure); 
\draw[->,line width=1*\zz pt]  (crew)  -- node [sloped,pos=0.24]{\sffamily\tiny{1.00 (1.00,0.00,0.00)}} (cabin); 
\draw[->,line width=1*\zz pt] (crew)  -- node [sloped,above]{\sffamily\tiny{1.00 (0.00,0.77,0.23)}} (booking); 
\draw[->,line width=1*\zz pt]  (crew)  -- node [sloped,pos=0.62]{\sffamily\tiny{1.00 (0.00,0.95,0.05)}} (checkin); 
\draw[->,line width=1*\zz pt] (crew)  -- node [sloped,above,pos=0.3]{\sffamily\tiny{1.00 (0.99,0.00,0.01)}} (meal); 
\draw[->,line width=1*\zz pt] (crew)  -- node [sloped]{\sffamily\tiny{1.00 (0.00,0.02,0.98)}} (overall); 
\draw[->,line width=1*\zz pt] (cabin)  -- node [sloped,pos=0.68]{\sffamily\tiny{1.00 (0.99,0.00,0.01)}} (meal); 
\draw[->,line width=1*\zz pt] (cabin)  -- node [sloped,pos=0.84]{\sffamily\tiny{1.00 (0.00,0.96,0.04)}} (booking); 
\draw[->,line width=1*\zz pt]  (cabin)  -- node [sloped]{\sffamily\tiny{1.00 (0.00,0.08,0.92)}} (departure); 
\draw[->,line width=0.96*\zz pt,black!96] (cabin)  -- node [sloped,pos=0.85]{\sffamily\tiny{0.96 (0.00,0.80,0.16)}} (checkin);
\draw[->,line width=0.455*\zz pt,black!45] (cabin)  -- node [sloped]{\sffamily\tiny{0.45 (0.00,0.03,0.42)}} (overall);  
\draw[->,line width=1.00*\zz pt]  (meal)  -- node [sloped,pos=.74,below]{\sffamily\tiny{1.00 (0.00,0.83,0.17)}} (booking);  
\draw[->,line width=1*\zz pt]  (meal)  -- node [sloped,below]{\sffamily\tiny{1.00 (0.01,0.21,0.78)}} (departure);  
 \draw[->,line width=0.04*\zz pt,black!4] (meal)  -- node [sloped,below]{\sffamily\tiny{0.04 (0.00,0.14,0.00)}} (checkin);  
\draw[->,line width=0.87*\zz pt,black!87] (meal)  -- node [sloped,below,pos=0.23]{\sffamily\tiny{0.87 (0.00,0.25,0.62)}} (overall);  
\draw[->,line width=1*\zz pt] (departure)  -- node [sloped,below,pos=0.15]{\sffamily\tiny{1.00 (0.00,0.63,0.37)}} (booking);  
 \draw[->,line width=1*\zz pt] (departure)  -- node [sloped,below,pos=0.27]{\sffamily\tiny{1.00 (0.00,0.58,0.42)}} (checkin);  
 \draw[->,line width=0.62*\zz pt,black!62]  (departure)  -- node [sloped,pos=0.25]{\sffamily\tiny{0.62 (0.00,0.18,0.44)}} (overall);  
\draw[->,line width=1*\zz pt,black] (booking)  -- node [sloped,below]{\sffamily\tiny{1.00 (0.00,0.02,0.92)}} (checkin);  
\draw[->,line width=0.785*\zz pt,black!78] (booking)  -- node [sloped,pos=0.25]{\sffamily\tiny{0.78 (0.00,0.36,0.42)}} (overall);  
\draw[->,line width=0.270*\zz pt,black!27] (checkin)  -- node [sloped,pos=0.25]{\sffamily\tiny{0.27 (0.00,0.21,0.06)}} (overall);  
\end{tikzpicture}
}
\end{center}
\caption{Edge strengths of the ALDAGs learned with the ST\_K4 algorithm over the airline dataset. In parenthesis, the proportion of times edges are given labels symmetric, context-specific, and local.}
\label{fig:aldag_K4}
\end{figure}

Figure \ref{fig:aldag_K4} reports the edge strengths for ALDAGs learned from 4-parents staged trees. The edges' width and color are proportional to their strength. It can be clearly seen that now the learned ALDAGs were not necessarily fully complete. Again, context-specific and local edges are more common than symmetric ones.

\begin{figure}
\begin{center}
\scalebox{0.9}{
  \begin{tikzpicture}[
        shorten >=1pt, auto, thick,
        node distance=3cm,
    main node/.style={circle,draw,font=\sffamily\tiny, minimum size = 1.2cm}
                            ]
      \node[main node] (crew) at (1*\xx,2.5*\yy) {Crew};
      \node[main node] (cabin) at (5*\xx,2.5*\yy) {Cabin};
	\node[main node] (departure) at (6*\xx,1*\yy) {Departure};
\node[main node] (meal) at (4*\xx,0*\yy) {Meal};
\node[main node] (checkin) at (2*\xx,0*\yy) {Check-in};
\node[main node] (booking) at (0*\xx,1*\yy){Booking};
\node[main node] (overall) at (3*\xx,3*\yy){Overall};
\draw[->,line width=1*\zz pt,black] (crew) -- (cabin); 
\draw[->,line width=1*\zz pt,green] (crew) -- (booking); 
\draw[->,line width=1*\zz pt,red] (crew) -- (checkin);
\draw[->,line width=1*\zz pt,green] (crew) -- (departure); 
\draw[->,line width=1*\zz pt,black] (crew) -- (meal);
\draw[->,line width=1*\zz pt,green] (crew) -- (overall);
\draw[->,line width=1*\zz pt,red] (cabin) -- (overall);   
\draw[->,line width=1*\zz pt,black] (cabin) -- (meal); 
\draw[->,line width=1*\zz pt,red] (cabin) -- (booking); 
\draw[->,line width=1*\zz pt,green] (cabin) -- (departure);
\draw[->,line width=1*\zz pt,red] (cabin) -- (checkin);  
\draw[->,line width=1*\zz pt,green] (meal) -- (departure); 
\draw[->,line width=1*\zz pt,green] (meal) -- (overall); 
\draw[->,line width=1*\zz pt,green] (meal) -- (booking);
\draw[->,line width=1*\zz pt,red] (departure) -- (checkin);
\draw[->,line width=1*\zz pt,red] (departure) -- (booking);   
\draw[->,line width=1*\zz pt,green] (departure) -- (overall);   
\draw[->,line width=1*\zz pt,green] (booking) -- (overall); 
\draw[->,line width=1*\zz pt,green] (booking) -- (checkin); 
\end{tikzpicture}
}
\end{center}
\caption{Averaged ALDAG learned from 4-parents staged trees over the airline dataset. \label{fig:average_airline}}
\end{figure}
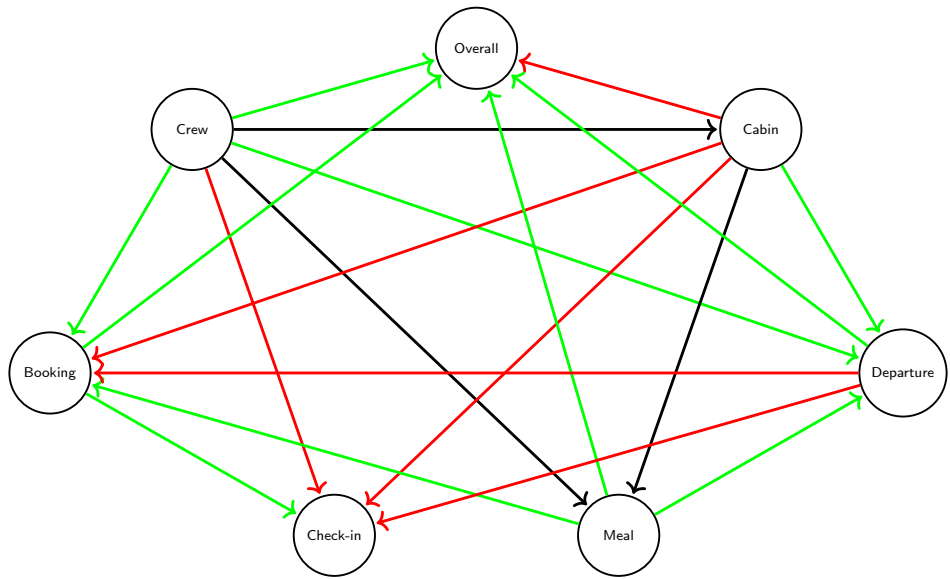

The ALDAG associated with the averaged staged tree $T^*$ learned with the algorithm for 4-parents staged trees is reported in Figure \ref{fig:average_airline}. It is not fully connected; it has two missing edges (Check-in, Overall) and (Meal, Check-in), three edges with symmetric labels, six with context-specific labels, and ten with local labels. The ALDAG intuitively shows that the check-in has no effect on the overall experience conditionally on all other service dimensions. It also shows that there are complex dependence patterns beyond symmetric ones.

\begin{figure}
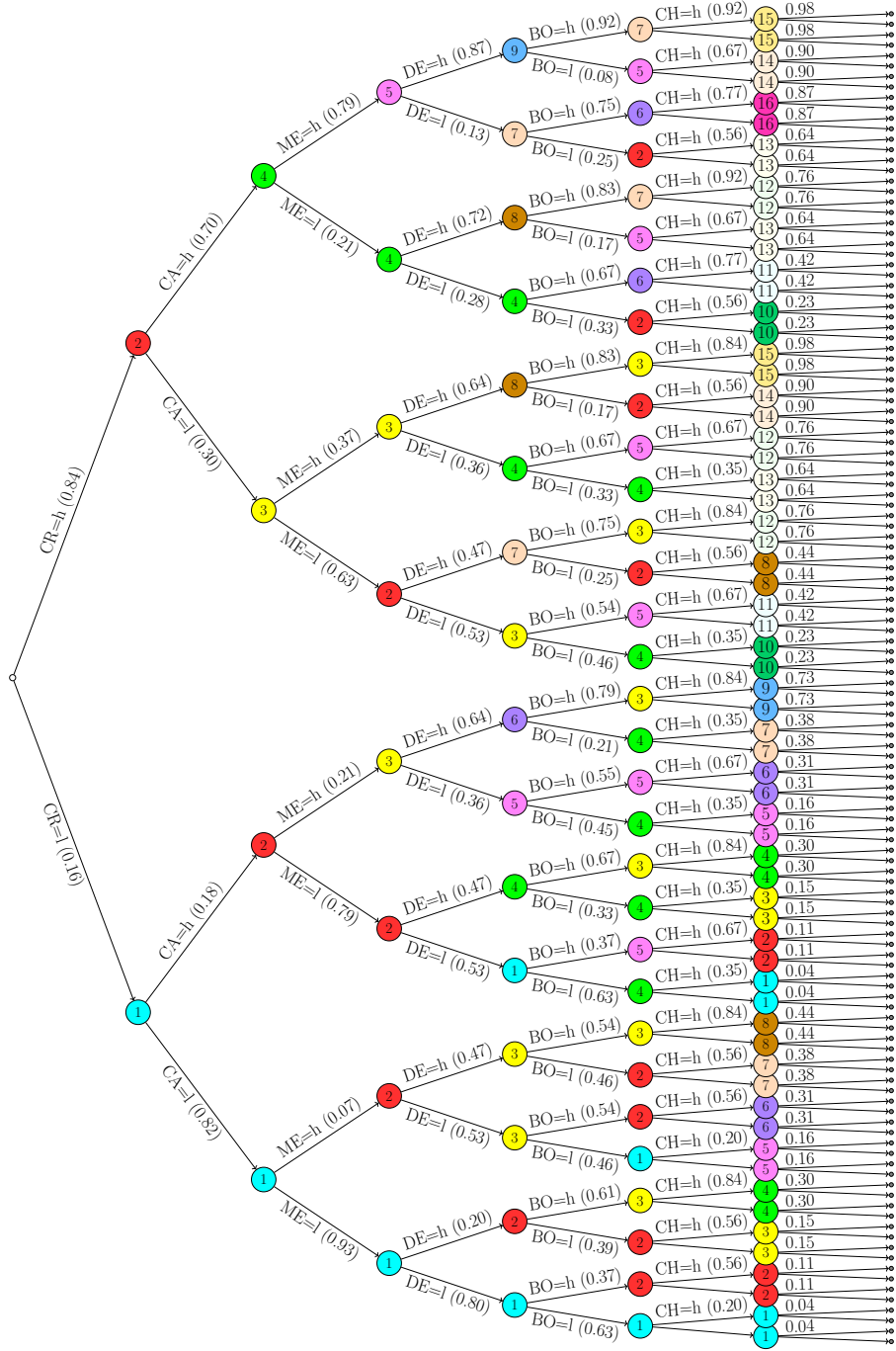

\begin{center}
\scalebox{0.275}{

}
\end{center}
\caption{Averaged staged tree using the 4-parents algorithms over the airline dataset. \label{fig:st_av}}
\end{figure}

\begin{figure}
    \centering
    \includegraphics[scale=0.4]{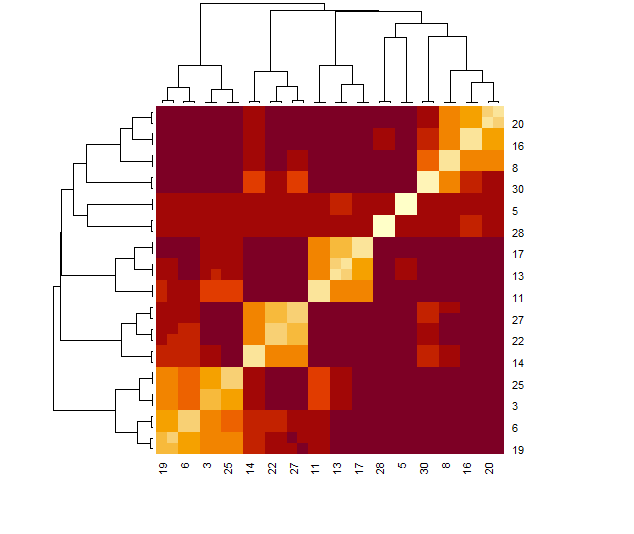}
    \caption{Heatmap of the pairwise dissimilarity matrix of the staging of the variable Check-in.}
    \label{fig:heatmap}
\end{figure}

Because the number of variables in the data is still limited, the averaged staged tree $T^*$ can still be visualized and is reported in Figure \ref{fig:st_av}\footnote{Because of the large size of the tree colors are repeated at different depths in the tree. However, staging should only be interpreted within vertices at the same depth.}. It shows a complex staging embedding non-symmetric dependence patterns, which can be identified by investigating the coloring of the non-leaf vertices. A complete interpretation of this tree is beyond the scope of this illustrative example and will, therefore, not be pursued. The staging of the averaged tree was chosen using the a priori fixed threshold of 0.5. However, the bootstrap approach allows for the investigation of a user's preferred staging using the heatmap of the pairwise dissimilarity matrix. As an illustration, Figure \ref{fig:heatmap} shows the heatmap for the staging of the variable Check-in quite strongly confirming the presence of seven stages as reported in the staged tree in Figure \ref{fig:st_av}.

\begin{figure}
    \centering
    \includegraphics[scale=0.45]{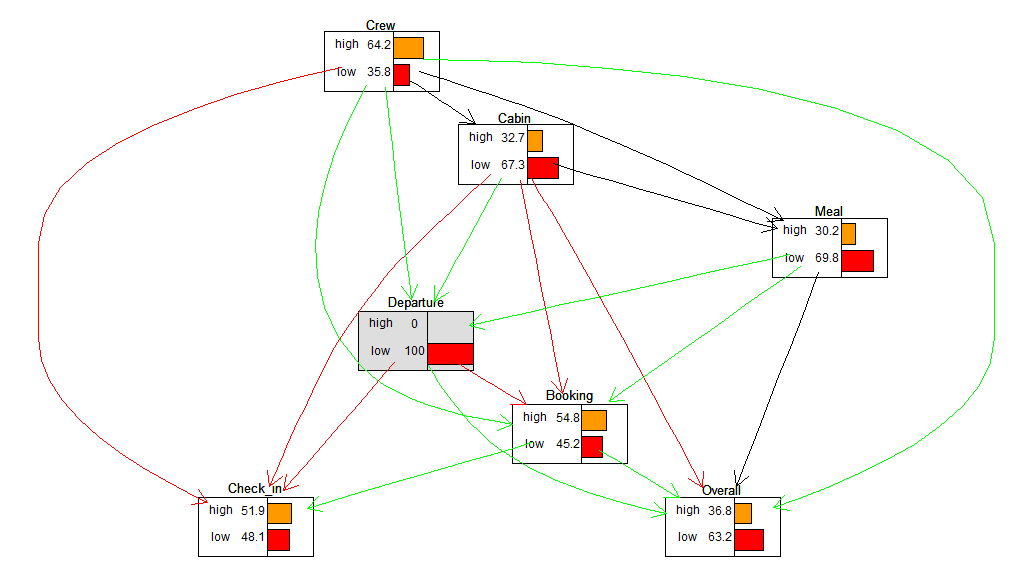}
    \caption{What-if sensitivity analysis using hard evidence for the airline passengers’ satisfaction.}
    \label{fig:hard_ev}
\end{figure}

Through the ALDAG, we can also perform what-if analyses using the fast propagation routines available for BNs and visualize the results concisely. The model estimates the marginal probability of a passenger being satisfied (high) as 70\%. As an illustration, suppose we are interested in assessing how this probability changes for passengers who are dissatisfied with the departure service. This is showcased in Figure \ref{fig:hard_ev} reporting the ALDAG together with introduced evidence (gray node) and the updated marginal probabilities. The probability of a satisfied passenger dramatically decreases to around 37\%. This type of what-if analysis is often said to be based on \textit{hard evidence} \citep{cugnata2016bayesian}.

\begin{figure}
    \centering
    \includegraphics[scale=0.45]{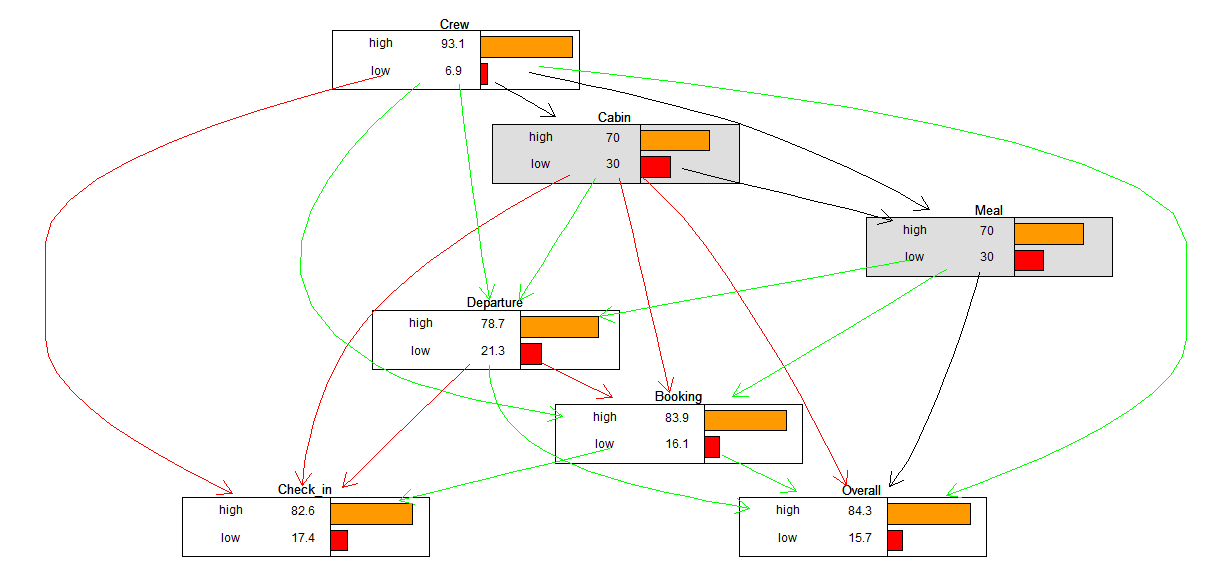}
    \caption{What-if sensitivity analysis using soft evidence for the airline passengers’ satisfaction.}
    \label{fig:soft_ev}
\end{figure}

As an additional illustration, the learned ALDAG states that the two service dimensions passengers are less satisfied with are the meal and the cabin (probability of 57\% and 61\% respectively). Consider passengers who are known to be slightly more likely to be satisfied with these two dimensions and that these probabilities are assumed to be equal to 70\%. Figure \ref{fig:soft_ev} reports this scenario and demonstrates that the probability of an overall satisfied passenger increased from 70\% to 84\%. In this case, the what-if analysis is said to be based on \textit{soft evidence}.

\subsection{Railway travelers’ satisfaction}
\label{sec:rail}

Our second application analyzes the satisfaction of railway travelers in the European Union. A similar analysis using BNs was developed in \citet{perucca2014travellers} using data collected between 2010 and 2014.

\subsubsection{The data}

The Flash Eurobarometer 463 \citep{ZA6933} collected information about Europeans’ satisfaction
with passenger rail service at the beginning of 2018 (this is the latest survey available about rail services). The Flash Eurobarometer were launched by the European Commission in the late eighties and are ``small scale"  surveys conducted in all EU member states at times, occasionally reducing or enlarging the scope of countries as a function of specific topics. The overall satisfaction of around 21k railway travelers across the European Union is recorded on a three-level scale (low/medium/high). Individual information about the surveyed travelers is also recorded: gender (male/female), age (dichotomized to 15-54/more than 54), education (dichotomized to university/no university), community (rural/small/large), occupation (self-employed/employee/unemployed), and country. Additional information about the location of the travelers was available for some countries at different spatial resolutions: national (e.g. Estonia and Croatia), NUTS1 (Nomenclature of territorial units for statistics) (e.g. Germany and Italy), or NUTS2 (e.g. Spain and Poland).\footnote{See \url{https://ec.europa.eu/eurostat/web/nuts/overview} for details about these geographic divisions.} This different geographical resolution was important when integrating additional data sources as discussed next.

As in \citet{perucca2014travellers}, we included three macroeconomic indicators, namely disposable income of households expressed in PPP (purchasing power parity), population density, and unemployment rates. Data was retrieved from the Eurostat Regional Database\footnote{From \url{https://ec.europa.eu/eurostat/databrowser/explore/all/all_themes}} and was available at the NUTS2 geographical resolution. Therefore, for countries whose data was available with less resolution, data aggregation and averaging were performed. These variables were included since the economic and social environment is known to have an effect on individuals' satisfaction \citep{fiorio}. The variables were then dichotomized using the quantile method into low/high.

We further included three variables characterizing the rail infrastructure in which the travel took place, as in \citet{perucca2014travellers}. The length of the railway (proportionally to the size of the region) was retrieved from the Eurostat Regional Database and measured at NUTS2 resolution. The demand for rail transport was measured by passenger-per-capita-kilometre and retrieved from the Independent Regulators' Group - Rail\footnote{Available at \url{https://irg-rail.eu/irg/documents/market-monitoring/260,2020.html}}. Rail fares were measured as
passenger revenue per passenger-kilometer, expressed in euros and converted to a common currency using purchasing parity exchange rates\footnote{Retrieved from the Organization for Economic Co-operation and Development (OECD) at \url{https://stats.oecd.org/index.aspx?DataSetCode=SNA_Table4}}. This variable was also retrieved from the Independent Regulators' Group - Rail and measured at the national level. Geographical resolution aggregations were also carried out for these variables, which were ultimately dichotomized into low/high.

Observations with missing values were dropped from the data. Ireland and the UK were not included in the study since some of the rail infrastructure and macroeconomic variables were not available.  The final dataset includes 20995 observations and 13 variables.

\begin{figure}
    \centering
    \includegraphics[scale=0.48]{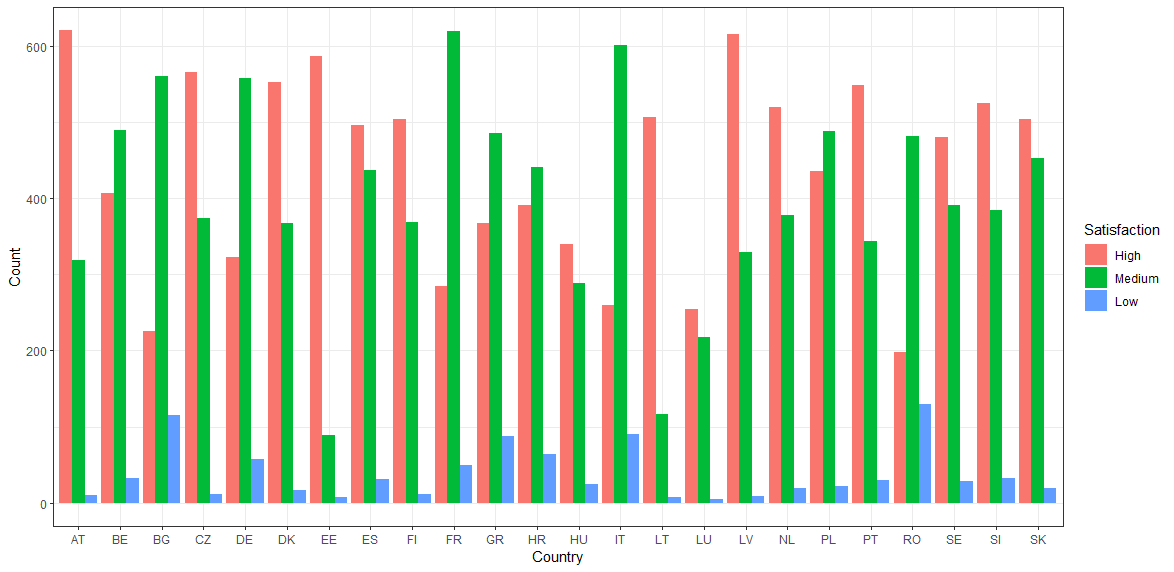}
    \caption{Barplot of the number of satisfied rail passengers by European country.}
    \label{fig:barplot}
\end{figure}

Figure \ref{fig:barplot} shows the satisfaction responses per country\footnote{ISO 3166-1 alpha-2 codes are used, see \url{https://en.wikipedia.org/wiki/List_of_ISO_3166_country_codes}.}, highlighting quite strong differences between countries. For instance, travelers in the Baltic countries are strongly satisfied with the service. On the other hand, travelers from Bulgaria and Romania are the most dissatisfied with the service. Because of these similarities, and to simplify the analysis with staged trees, which heavily depends on the sample space size, we categorized countries into four regions (Eastern, Western, Southern, and Northern Europe)\footnote{Using the division from \url{https://en.wikipedia.org/wiki/Regions_of_Europe}.}

Instead of visualizing the relationship between each predictor and satisfaction via barplots as in Figure \ref{fig:barplot}, we use PGMs to integrate all data sources and study the joint effect of all predictors, as well as their interdependencies, on the travelers' satisfaction.

\subsubsection{Model selection}
Model selection was performed in a similar way as in Section \ref{sec:air}. However, the large number of variables considered made impossible the investigation of all possible orders. For this reason, we grouped variables into three groups: demographics, rail-related, and macroeconomic. An optimal order among the variables in each group was identified using resampling. Once these were identified, an optimal order of the three groups was found using resampling again.

\begin{figure}
    \centering
    \includegraphics[scale=0.5]{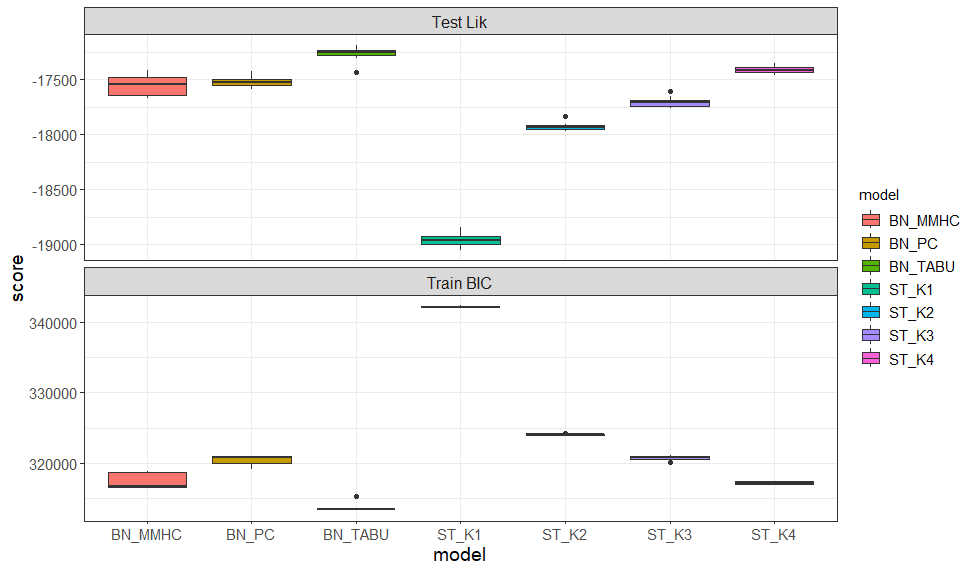}
    \caption{Boxplots of BIC scores (train dataset) and log-likelihood (test dataset) from a 10-fold
cross-validation over the railway dataset.}
    \label{fig:cross}
\end{figure}

All algorithms considered in Section \ref{sec:air} were investigated, with the exception of BHC which would not be feasible with this larger number of variables. The results of the cross-validation are reported in Figure \ref{fig:cross}. Compared to the airline application, BNs learned with the tabu algorithm are the best scoring model. Staged trees learned with the algorithm for 4-parents staged trees are the second-best scoring approach. This might be due to the fact that a higher number of parents might have had to be considered. The average number of parameters of the learned BNs with tabu is 324, while for staged trees it is only 232. Adding an additional parent might have provided the best scoring model, but this avenue was not pursued to avoid having an overly complex ALDAG. For illustrative purposes, we select the staged tree obtained with the 4-parents routine as our favorite one.

\begin{figure}
    \centering
    \includegraphics[scale=0.65]{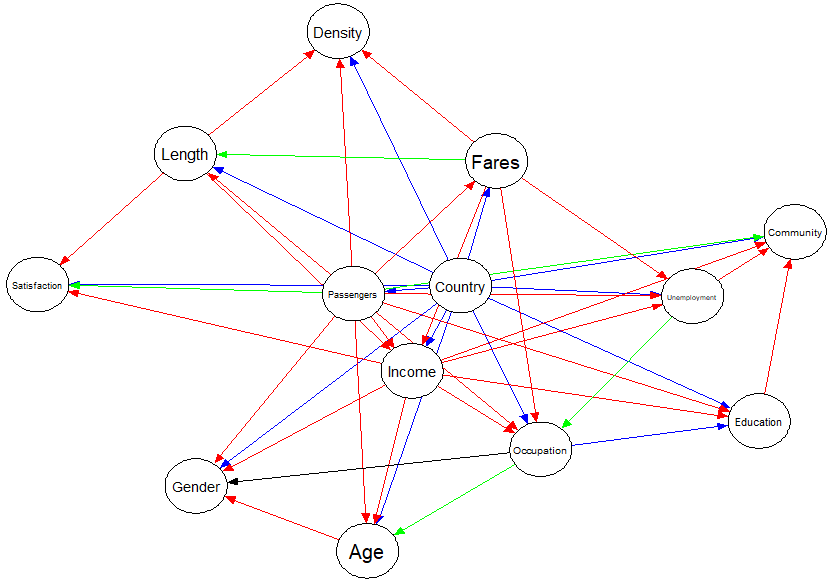}
    \caption{ALDAG associated to the averaged staged tree learned using 4-parents staged tree algorithm over the railway dataset.}
    \label{fig:aldag4}
\end{figure}

Because of the larger number of variables the averaged staged tree cannot be fully visualized: it would include more than 50k root-to-leaf paths. Figure \ref{fig:aldag4} reports the ALDAG associated with the averaged staged tree. There are 45 edges, of which one has label symmetric (black), 26 have label context-specific (red), 13 have label partial (blue), and 5 have label local (green). The maximum in-degree is five, so that sparsity is still retained even when using resampling techniques. The ALDAG shows a great level of interdependence between the predictors of satisfaction, which would be, in general, overlooked by ``traditional" methods such as logistic regression.

\subsection{Model interpretation}

The ALDAG in Figure \ref{fig:aldag4} shows that railway length, country, income of households, and the demand/proportion of passengers have a direct effect on service satisfaction; all other predictors are conditionally independent of satisfaction. This confirms the conclusions of \citet{perucca2014travellers}, who observed that country has a direct effect on satisfaction, while all other demographic information is only relatively informative. 

\begin{figure}
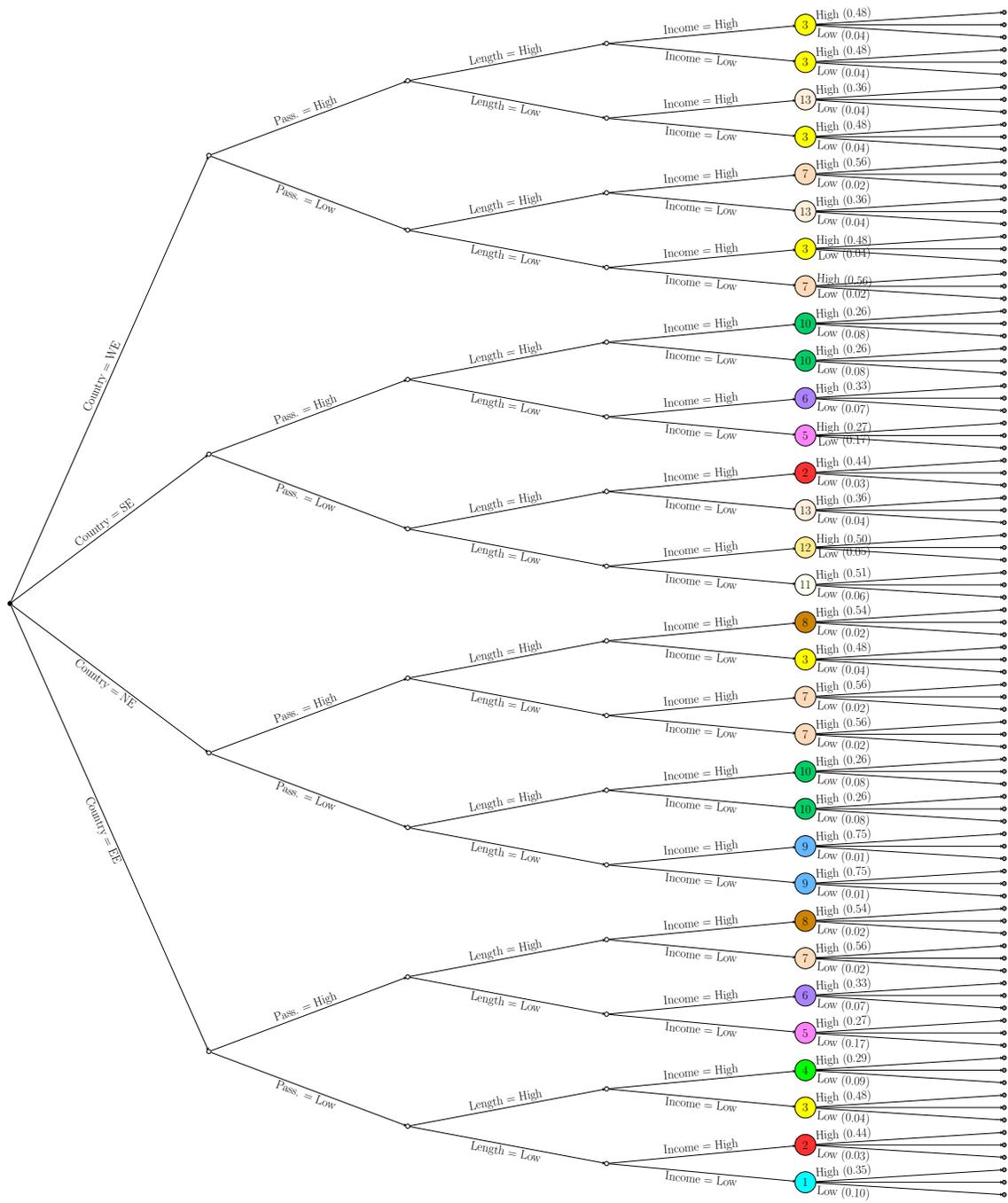

\begin{center}
\scalebox{0.21}{
    %
}
\end{center}
\caption{Dependence subtree associated to the variable Satisfaction in the ALDAG in Figure \ref{fig:aldag4}. \label{fig:st_av1}}
\end{figure}

The actual relationship between satisfaction and its parents can be visualized using the dependence subtree, reported in Figure \ref{fig:st_av1}. Again, a complete interpretation of this tree is beyond the scope of this paper, but a few interesting patterns can be mentioned. The stage with the highest probability of low satisfaction is number five, corresponding to passengers coming from regions with a high demand/passenger proportion, low track length, and low income from Southern and Eastern Europe. On the other hand, passengers with the highest probability of high satisfaction come from regions of Northern Europe with low demand and length, irrespective of household income (Stage 9). 

\begin{table}[]
    \centering
    \scalebox{0.8}{
\begin{tabular}{l|ccc|c}
    \toprule
    & \multicolumn{3}{c|}{Satisfaction} & \\
    Predictor & Low & Medium & High & Mutual Info\\
    \midrule
    Country (SE/EE/WE/NE) &\textcolor{red}{0.046} &\textcolor{black}{0.160} & \textcolor{blue}{0.202} & 0.015996\\
      Gender (Male/Female) &\textcolor{blue}{0.001} &\textcolor{red}{0.001} & \textcolor{blue}{0.000} & 0.000002\\
       Age (15-54/more than 54) &\textcolor{red}{0.002} &\textcolor{blue}{0.006} & \textcolor{red}{0.004} & 0.000029\\
       Education (no university/university) &\textcolor{red}{0.004} &\textcolor{red}{0.010} & \textcolor{blue}{0.014} & 0.000116\\
        Community (rural/small/large) &\textcolor{black}{0.001} &\textcolor{red}{0.018} & \textcolor{blue}{0.018} & 0.000127\\
             Occupation (not working/self-employed/employee) &\textcolor{black}{0.002} &\textcolor{red}{0.009} & \textcolor{blue}{0.011} & 0.000060\\
    Fares (Low/High) &\textcolor{red}{0.012} &\textcolor{blue}{0.035} & \textcolor{red}{0.024} & 0.000873 \\
        Passengers (Low/High) &\textcolor{red}{0.016} &\textcolor{blue}{0.038} & \textcolor{red}{0.022} & 0.001259\\
    Length (Low/High) &\textcolor{red}{0.014} &\textcolor{blue}{0.054} & \textcolor{red}{0.041} & 0.001788\\
    Unemployment (Low/High) &\textcolor{blue}{0.007} & \textcolor{red}{0.005}& \textcolor{red}{0.002}& 0.000156\\
    Density (Low/High) & \textcolor{red}{0.006} & \textcolor{blue}{0.054}& \textcolor{red}{0.048} &  0.001474\\
         Income (Low/High)& \textcolor{red}{0.015} & \textcolor{blue}{0.042} & \textcolor{red}{0.027} & 0.001289\\
\bottomrule
    \end{tabular}}
    \caption{Maximum change in Satisfaction probabilities computed by What-if analyses using each predictor. The color represents the direction of the change. The last column reports the mutual information between each predictor and Satisfaction}
    \label{tab:my_label}
\end{table}

We then performed an extensive what-if analysis to assess the effect of each predictor on satisfaction. Table \ref{tab:my_label} reports the maximum absolute change in the probability that satisfaction is equal to a specific value when each predictor is fixed. The color describes the direction of the change: red-decrease, blue-increase, black-no uniform pattern. It can be seen that changes are small, with the exception of the variable Country. By changing country from Southern Europe to Northern Europe the probability of a highly satisfied traveler increases by 0.202. By changing the value of Fares, Passengers, Length, Density, and Income from Low to High, the probability of Low or High Satisfaction decreases, while the probability of a Medimum Satisfaction increases.

Table \ref{tab:my_label} further reports the mutual information between the output variable Satisfaction and each of the predictors computed from the model. Mutual information is a standard sensitivity measure to assess the strength of relationship in PGMs \citep[e.g.][]{kjaerulff2008bayesian}. Again, Country has clearly the strongest effect on Satisfaction, followed by Length and Density.  

\section{Discussion}

Staged trees have proven to be a powerful PGM to describe complex patterns of dependence in tabular data. Furthermore, the associated ALDAG and dependence subtrees provide an intuitive graphical representation to visualize these complex patterns for larger applications such as the one in railways evaluation investigated here. As a PGM, staged trees are naturally suited for the integration of heterogeneous data sources, since each variable of the associated ALDAG can be informed by its own data source.

Because of the complexity and size of the model search space, it has been noticed that staged trees tend to overfit the training data and have lower performance over test sets.  In this paper, we have provided a solution to this problem by introducing robust modeling approaches based on data resampling and cross-validation, which can be applied to learn both the variables' ordering in the tree and the staging of the vertices once the order is fixed. These methods were implemented using the \texttt{stagedtrees} R package and we plan to include them in the next release of the package on CRAN.

The two data analyses showcased the applied use of these routines and the insights staged trees, coupled with their ALDAG representation, can provide in practice.  Sensitivity methods and what-if analyses that are standard in BNs have been used for the first time in staged tree models by taking advantage of the underlying ALDAG. We plan to also include such capabilities in the next release of \texttt{stagedtrees}.

Just as for BNs, an alternative approach for robust learning of staged trees would be to take a fully Bayesian approach and use MCMC algorithms to create a posterior sample of staged trees. Bayesian clustering methods could be almost directly applied to the Bayesian structural learning of staged trees, coupled with standard clustering averaging methods to identify a stage tree point estimate from the sample \citep{wade2023bayesian}.  

One drawback of the approach proposed here is that in order to impose sparsity we used algorithms to learn $k$-parents staged trees, but their averaged estimate does not necessarily fall within the same class of models. This issue could be avoided using the above-mentioned Bayesian approach by defining appropriate prior distributions and sampling schemes that would only explore models within the required class. The development of this approach is the focus of current research and will be reported in the near future.

\section*{Acknowledgments}
The authors gratefully acknowledge the help from Prof. Silvia Salini who provided the data of the airline application in Section \ref{sec:air} and the code used to produce Figures \ref{fig:hard_ev} and \ref{fig:soft_ev}.

\bibliographystyle{elsarticle-harv} 
\bibliography{bib}

\begin{thebibliography}{79}
\expandafter\ifx\csname natexlab\endcsname\relax\def\natexlab#1{#1}\fi
\providecommand{\url}[1]{\texttt{#1}}
\providecommand{\href}[2]{#2}
\providecommand{\path}[1]{#1}
\providecommand{\DOIprefix}{doi:}
\providecommand{\ArXivprefix}{arXiv:}
\providecommand{\URLprefix}{URL: }
\providecommand{\Pubmedprefix}{pmid:}
\providecommand{\doi}[1]{\href{http://dx.doi.org/#1}{\path{#1}}}
\providecommand{\Pubmed}[1]{\href{pmid:#1}{\path{#1}}}
\providecommand{\bibinfo}[2]{#2}
\ifx\xfnm\relax \def\xfnm[#1]{\unskip,\space#1}\fi
\bibitem[{Barclay et~al.(2014)Barclay, Hutton and Smith}]{barclay2014chain}
\bibinfo{author}{Barclay, L.M.}, \bibinfo{author}{Hutton, J.},
  \bibinfo{author}{Smith, J.}, \bibinfo{year}{2014}.
\newblock \bibinfo{title}{Chain event graphs for informed missingness}.
\newblock \bibinfo{journal}{Bayesian Analysis} \bibinfo{volume}{9},
  \bibinfo{pages}{53--76}.
\bibitem[{Borgonovo(2023)}]{borgonovo2023sensitivity}
\bibinfo{author}{Borgonovo, E.}, \bibinfo{year}{2023}.
\newblock \bibinfo{title}{Sensitivity analysis}.
\newblock \bibinfo{journal}{Tutorials in Operations Research: Advancing the
  Frontiers of OR/MS: From Methodologies to Applications} ,
  \bibinfo{pages}{52--81}.
\bibitem[{Boutilier et~al.(1996)Boutilier, Friedman, Goldszmidt and
  Koller}]{Boutilier1996}
\bibinfo{author}{Boutilier, C.}, \bibinfo{author}{Friedman, N.},
  \bibinfo{author}{Goldszmidt, M.}, \bibinfo{author}{Koller, D.},
  \bibinfo{year}{1996}.
\newblock \bibinfo{title}{{Context-specific independence in Bayesian
  networks}}, in: \bibinfo{booktitle}{Proceedings of the 12th Conference on
  Uncertainty in Artificial Intelligence}, pp. \bibinfo{pages}{115--123}.
\bibitem[{Briganti et~al.(2022)Briganti, Decety, Scutari, McNally and
  Linkowski}]{briganti2022using}
\bibinfo{author}{Briganti, G.}, \bibinfo{author}{Decety, J.},
  \bibinfo{author}{Scutari, M.}, \bibinfo{author}{McNally, R.J.},
  \bibinfo{author}{Linkowski, P.}, \bibinfo{year}{2022}.
\newblock \bibinfo{title}{Using {B}ayesian networks to investigate
  psychological constructs: The case of empathy}.
\newblock \bibinfo{journal}{Psychological Reports} ,
  \bibinfo{pages}{00332941221146711}.
\bibitem[{Caravagna and Ramazzotti(2021)}]{caravagna2021learning}
\bibinfo{author}{Caravagna, G.}, \bibinfo{author}{Ramazzotti, D.},
  \bibinfo{year}{2021}.
\newblock \bibinfo{title}{Learning the structure of {B}ayesian networks via the
  bootstrap}.
\newblock \bibinfo{journal}{Neurocomputing} \bibinfo{volume}{448},
  \bibinfo{pages}{48--59}.
\bibitem[{Carli et~al.(2022)Carli, Leonelli, Riccomagno and
  Varando}]{Carli2022}
\bibinfo{author}{Carli, F.}, \bibinfo{author}{Leonelli, M.},
  \bibinfo{author}{Riccomagno, E.}, \bibinfo{author}{Varando, G.},
  \bibinfo{year}{2022}.
\newblock \bibinfo{title}{The {R} package stagedtrees for structural learning
  of stratified staged trees}.
\newblock \bibinfo{journal}{Journal of Statistical Software}
  \bibinfo{volume}{102}, \bibinfo{pages}{1--30}.
\bibitem[{Carli et~al.(2023)Carli, Leonelli and Varando}]{carli2020new}
\bibinfo{author}{Carli, F.}, \bibinfo{author}{Leonelli, M.},
  \bibinfo{author}{Varando, G.}, \bibinfo{year}{2023}.
\newblock \bibinfo{title}{A new class of generative classifiers based on staged
  tree models}.
\newblock \bibinfo{journal}{Knowledge-Based Systems} , \bibinfo{pages}{110488}.
\bibitem[{Castelletti and Peluso(2021)}]{castelletti2021equivalence}
\bibinfo{author}{Castelletti, F.}, \bibinfo{author}{Peluso, S.},
  \bibinfo{year}{2021}.
\newblock \bibinfo{title}{Equivalence class selection of categorical graphical
  models}.
\newblock \bibinfo{journal}{Computational Statistics \& Data Analysis}
  \bibinfo{volume}{164}, \bibinfo{pages}{107304}.
\bibitem[{Ceriani and Gigliarano(2020)}]{ceriani2020multidimensional}
\bibinfo{author}{Ceriani, L.}, \bibinfo{author}{Gigliarano, C.},
  \bibinfo{year}{2020}.
\newblock \bibinfo{title}{Multidimensional well-being: {A} {B}ayesian networks
  approach}.
\newblock \bibinfo{journal}{Social Indicators Research} \bibinfo{volume}{152},
  \bibinfo{pages}{237--263}.
\bibitem[{Chickering et~al.(1997)Chickering, Heckerman and
  Meek}]{chickering1997bayesian}
\bibinfo{author}{Chickering, D.M.}, \bibinfo{author}{Heckerman, D.},
  \bibinfo{author}{Meek, C.}, \bibinfo{year}{1997}.
\newblock \bibinfo{title}{A {B}ayesian approach to learning {B}ayesian networks
  with local structure}, in: \bibinfo{booktitle}{Proceedings of the 13th
  Conference on Uncertainty in Artificial Intelligence}, pp.
  \bibinfo{pages}{80--89}.
\bibitem[{Collazo et~al.(2018)Collazo, G{\"o}rgen and Smith}]{collazo2018chain}
\bibinfo{author}{Collazo, R.A.}, \bibinfo{author}{G{\"o}rgen, C.},
  \bibinfo{author}{Smith, J.Q.}, \bibinfo{year}{2018}.
\newblock \bibinfo{title}{Chain event graphs}.
\newblock \bibinfo{publisher}{CRC Press}.
\bibitem[{Cowell and Smith(2014)}]{cowell2014causal}
\bibinfo{author}{Cowell, R.}, \bibinfo{author}{Smith, J.},
  \bibinfo{year}{2014}.
\newblock \bibinfo{title}{Causal discovery through {MAP} selection of
  stratified chain event graphs}.
\newblock \bibinfo{journal}{Electronic Journal of Statistics}
  \bibinfo{volume}{8}, \bibinfo{pages}{965--997}.
\bibitem[{Cugnata et~al.(2014)Cugnata, Kenett and Salini}]{cugnata2014bayesian}
\bibinfo{author}{Cugnata, F.}, \bibinfo{author}{Kenett, R.},
  \bibinfo{author}{Salini, S.}, \bibinfo{year}{2014}.
\newblock \bibinfo{title}{Bayesian network applications to customer surveys and
  infoq}.
\newblock \bibinfo{journal}{Procedia Economics and Finance}
  \bibinfo{volume}{17}, \bibinfo{pages}{3--9}.
\bibitem[{Cugnata et~al.(2016)Cugnata, Kenett and Salini}]{cugnata2016bayesian}
\bibinfo{author}{Cugnata, F.}, \bibinfo{author}{Kenett, R.S.},
  \bibinfo{author}{Salini, S.}, \bibinfo{year}{2016}.
\newblock \bibinfo{title}{Bayesian networks in survey data: Robustness and
  sensitivity issues}.
\newblock \bibinfo{journal}{Journal of Quality Technology}
  \bibinfo{volume}{48}, \bibinfo{pages}{253--264}.
\bibitem[{Dahl et~al.(2022)Dahl, Johnson and M{\"u}ller}]{dahl2022search}
\bibinfo{author}{Dahl, D.B.}, \bibinfo{author}{Johnson, D.J.},
  \bibinfo{author}{M{\"u}ller, P.}, \bibinfo{year}{2022}.
\newblock \bibinfo{title}{Search algorithms and loss functions for {B}ayesian
  clustering}.
\newblock \bibinfo{journal}{Journal of Computational and Graphical Statistics}
  \bibinfo{volume}{31}, \bibinfo{pages}{1189--1201}.
\bibitem[{Dawid(1979)}]{dawid1979conditional}
\bibinfo{author}{Dawid, A.P.}, \bibinfo{year}{1979}.
\newblock \bibinfo{title}{Conditional independence in statistical theory}.
\newblock \bibinfo{journal}{Journal of the Royal Statistical Society Series B}
  \bibinfo{volume}{41}, \bibinfo{pages}{1--15}.
\bibitem[{Di~Pietro et~al.(2017)Di~Pietro, Mugion, Musella, Renzi and
  Vicard}]{di2017monitoring}
\bibinfo{author}{Di~Pietro, L.}, \bibinfo{author}{Mugion, R.G.},
  \bibinfo{author}{Musella, F.}, \bibinfo{author}{Renzi, M.F.},
  \bibinfo{author}{Vicard, P.}, \bibinfo{year}{2017}.
\newblock \bibinfo{title}{Monitoring an airport check-in process by using
  {B}ayesian networks}.
\newblock \bibinfo{journal}{Transportation Research Part A: Policy and
  Practice} \bibinfo{volume}{106}, \bibinfo{pages}{235--247}.
\bibitem[{D{\'\i}ez-Mesa et~al.(2018)D{\'\i}ez-Mesa, de~O{\~n}a and
  de~O{\~n}a}]{diez2018bayesian}
\bibinfo{author}{D{\'\i}ez-Mesa, F.}, \bibinfo{author}{de~O{\~n}a, R.},
  \bibinfo{author}{de~O{\~n}a, J.}, \bibinfo{year}{2018}.
\newblock \bibinfo{title}{{Bayesian networks and structural equation modelling
  to develop service quality models: Metro of {S}eville case study}}.
\newblock \bibinfo{journal}{Transportation Research Part A: Policy and
  Practice} \bibinfo{volume}{118}, \bibinfo{pages}{1--13}.
\bibitem[{Duarte and Solus(2021)}]{duarte2021representation}
\bibinfo{author}{Duarte, E.}, \bibinfo{author}{Solus, L.},
  \bibinfo{year}{2021}.
\newblock \bibinfo{title}{Representation of context-specific causal models with
  observational and interventional data}.
\newblock \bibinfo{journal}{arXiv:2101.09271} .
\bibitem[{Eggeling et~al.(2019)Eggeling, Grosse and
  Koivisto}]{eggeling2019algorithms}
\bibinfo{author}{Eggeling, R.}, \bibinfo{author}{Grosse, I.},
  \bibinfo{author}{Koivisto, M.}, \bibinfo{year}{2019}.
\newblock \bibinfo{title}{Algorithms for learning parsimonious context trees}.
\newblock \bibinfo{journal}{Machine Learning} \bibinfo{volume}{108},
  \bibinfo{pages}{879--911}.
\bibitem[{European~Commission(2018)}]{ZA6933}
\bibinfo{author}{European~Commission, B.}, \bibinfo{year}{2018}.
\newblock \bibinfo{title}{Flash {E}urobarometer 463 ({E}uropeans’
  satisfaction with passenger rail services)}.
\newblock \bibinfo{howpublished}{GESIS Datenarchiv, K{\"o}ln. ZA6933 Datenfile
  Version 1.0.0, https://doi.org/10.4232/1.13149}.
\newblock \DOIprefix\doi{10.4232/1.13149}.
\bibitem[{Fiorio et~al.(2007)Fiorio, Florio, Salini and Ferrari}]{fiorio}
\bibinfo{author}{Fiorio, C.V.}, \bibinfo{author}{Florio, M.},
  \bibinfo{author}{Salini, S.}, \bibinfo{author}{Ferrari, P.},
  \bibinfo{year}{2007}.
\newblock \bibinfo{title}{{Consumers' Attitudes on Services of General Interest
  in the {EU}: Accessibility, Price and Quality 2000-2004}}.
\newblock \bibinfo{type}{Privatisation Regulation Corporate Governance Working
  Papers} \bibinfo{number}{12195}. Fondazione Eni Enrico Mattei (FEEM).
\bibitem[{Freeman and Smith(2011)}]{freeman2011bayesian}
\bibinfo{author}{Freeman, G.}, \bibinfo{author}{Smith, J.Q.},
  \bibinfo{year}{2011}.
\newblock \bibinfo{title}{Bayesian {MAP} model selection of chain event
  graphs}.
\newblock \bibinfo{journal}{Journal of Multivariate Analysis}
  \bibinfo{volume}{102}, \bibinfo{pages}{1152--1165}.
\bibitem[{Friedman and Goldszmidt(1996)}]{friedman1996learning}
\bibinfo{author}{Friedman, N.}, \bibinfo{author}{Goldszmidt, M.},
  \bibinfo{year}{1996}.
\newblock \bibinfo{title}{Learning {B}ayesian networks with local structure},
  in: \bibinfo{booktitle}{Proceedings of the 12th Conference on Uncertainty in
  Artificial Intelligence}, pp. \bibinfo{pages}{252--262}.
\bibitem[{Friedman et~al.(1999)Friedman, Goldszmidt and
  Wyner}]{friedman1999data}
\bibinfo{author}{Friedman, N.}, \bibinfo{author}{Goldszmidt, M.},
  \bibinfo{author}{Wyner, A.}, \bibinfo{year}{1999}.
\newblock \bibinfo{title}{Data analysis with {B}ayesian networks: A bootstrap
  approach}, in: \bibinfo{booktitle}{Proceedings of the 15th Conference on
  Uncertainty in Artificial Intelligence}, pp. \bibinfo{pages}{196--205}.
\bibitem[{Friedman and Koller(2003)}]{friedman2003being}
\bibinfo{author}{Friedman, N.}, \bibinfo{author}{Koller, D.},
  \bibinfo{year}{2003}.
\newblock \bibinfo{title}{Being {B}ayesian about network structure. {A}
  {B}ayesian approach to structure discovery in {B}ayesian networks}.
\newblock \bibinfo{journal}{Machine Learning} \bibinfo{volume}{50},
  \bibinfo{pages}{95--125}.
\bibitem[{Glymour et~al.(2019)Glymour, Zhang and Spirtes}]{glymour2019review}
\bibinfo{author}{Glymour, C.}, \bibinfo{author}{Zhang, K.},
  \bibinfo{author}{Spirtes, P.}, \bibinfo{year}{2019}.
\newblock \bibinfo{title}{Review of causal discovery methods based on graphical
  models}.
\newblock \bibinfo{journal}{Frontiers in Genetics} \bibinfo{volume}{10},
  \bibinfo{pages}{524}.
\bibitem[{G{\"o}rgen et~al.(2018)G{\"o}rgen, Bigatti, Riccomagno and
  Smith}]{gorgen2018discovery}
\bibinfo{author}{G{\"o}rgen, C.}, \bibinfo{author}{Bigatti, A.},
  \bibinfo{author}{Riccomagno, E.}, \bibinfo{author}{Smith, J.Q.},
  \bibinfo{year}{2018}.
\newblock \bibinfo{title}{Discovery of statistical equivalence classes using
  computer algebra}.
\newblock \bibinfo{journal}{International Journal of Approximate Reasoning}
  \bibinfo{volume}{95}, \bibinfo{pages}{167--184}.
\bibitem[{G{\"o}rgen et~al.(2022)G{\"o}rgen, Leonelli and
  Marigliano}]{gorgen2022curved}
\bibinfo{author}{G{\"o}rgen, C.}, \bibinfo{author}{Leonelli, M.},
  \bibinfo{author}{Marigliano, O.}, \bibinfo{year}{2022}.
\newblock \bibinfo{title}{The curved exponential family of a staged tree}.
\newblock \bibinfo{journal}{Electronic Journal of Statistics}
  \bibinfo{volume}{16}, \bibinfo{pages}{2607--2620}.
\bibitem[{G{\"o}rgen et~al.(2015)G{\"o}rgen, Leonelli and
  Smith}]{gorgen2015differential}
\bibinfo{author}{G{\"o}rgen, C.}, \bibinfo{author}{Leonelli, M.},
  \bibinfo{author}{Smith, J.Q.}, \bibinfo{year}{2015}.
\newblock \bibinfo{title}{A differential approach for staged trees}, in:
  \bibinfo{booktitle}{Symbolic and Quantitative Approaches to Reasoning with
  Uncertainty}, pp. \bibinfo{pages}{346--355}.
\bibitem[{Goudie and Mukherjee(2016)}]{goudie2016gibbs}
\bibinfo{author}{Goudie, R.J.}, \bibinfo{author}{Mukherjee, S.},
  \bibinfo{year}{2016}.
\newblock \bibinfo{title}{A {G}ibbs sampler for learning {DAGs}}.
\newblock \bibinfo{journal}{The Journal of Machine Learning Research}
  \bibinfo{volume}{17}, \bibinfo{pages}{1032--1070}.
\bibitem[{Hua et~al.(2021)Hua, Feng, Ding and Ruan}]{hua2021bayesian}
\bibinfo{author}{Hua, W.}, \bibinfo{author}{Feng, X.}, \bibinfo{author}{Ding,
  C.}, \bibinfo{author}{Ruan, Z.}, \bibinfo{year}{2021}.
\newblock \bibinfo{title}{Bayesian network modeling analyzes of perceived urban
  rail transfer time}.
\newblock \bibinfo{journal}{Transportation Letters} \bibinfo{volume}{13},
  \bibinfo{pages}{514--521}.
\bibitem[{Jaeger et~al.(2006)Jaeger, Nielsen and Silander}]{jaeger2006learning}
\bibinfo{author}{Jaeger, M.}, \bibinfo{author}{Nielsen, J.D.},
  \bibinfo{author}{Silander, T.}, \bibinfo{year}{2006}.
\newblock \bibinfo{title}{Learning probabilistic decision graphs}.
\newblock \bibinfo{journal}{International Journal of Approximate Reasoning}
  \bibinfo{volume}{42}, \bibinfo{pages}{84--100}.
\bibitem[{Johnson and Mengersen(2012)}]{johnson2012integrated}
\bibinfo{author}{Johnson, S.}, \bibinfo{author}{Mengersen, K.},
  \bibinfo{year}{2012}.
\newblock \bibinfo{title}{Integrated {B}ayesian network framework for modeling
  complex ecological issues}.
\newblock \bibinfo{journal}{Integrated Environmental Assessment and Management}
  \bibinfo{volume}{8}, \bibinfo{pages}{480--490}.
\bibitem[{Kitson et~al.(2023)Kitson, Constantinou, Guo, Liu and
  Chobtham}]{kitson2023survey}
\bibinfo{author}{Kitson, N.K.}, \bibinfo{author}{Constantinou, A.C.},
  \bibinfo{author}{Guo, Z.}, \bibinfo{author}{Liu, Y.},
  \bibinfo{author}{Chobtham, K.}, \bibinfo{year}{2023}.
\newblock \bibinfo{title}{A survey of {B}ayesian network structure learning}.
\newblock \bibinfo{journal}{Artificial Intelligence Review} ,
  \bibinfo{pages}{1--94}.
\bibitem[{Kjaerulff and Madsen(2008)}]{kjaerulff2008bayesian}
\bibinfo{author}{Kjaerulff, U.B.}, \bibinfo{author}{Madsen, A.L.},
  \bibinfo{year}{2008}.
\newblock \bibinfo{title}{Bayesian networks and influence diagrams}.
\newblock \bibinfo{journal}{Springer Science+ Business Media}
  \bibinfo{volume}{200}, \bibinfo{pages}{114}.
\bibitem[{Koller and Friedman(2009)}]{koller2009probabilistic}
\bibinfo{author}{Koller, D.}, \bibinfo{author}{Friedman, N.},
  \bibinfo{year}{2009}.
\newblock \bibinfo{title}{Probabilistic graphical models: Principles and
  techniques}.
\newblock \bibinfo{publisher}{MIT Press}.
\bibitem[{Kuipers and Moffa(2017)}]{kuipers2017partition}
\bibinfo{author}{Kuipers, J.}, \bibinfo{author}{Moffa, G.},
  \bibinfo{year}{2017}.
\newblock \bibinfo{title}{Partition {MCMC} for inference on acyclic digraphs}.
\newblock \bibinfo{journal}{Journal of the American Statistical Association}
  \bibinfo{volume}{112}, \bibinfo{pages}{282--299}.
\bibitem[{Kuipers et~al.(2022)Kuipers, Suter and Moffa}]{kuipers2022efficient}
\bibinfo{author}{Kuipers, J.}, \bibinfo{author}{Suter, P.},
  \bibinfo{author}{Moffa, G.}, \bibinfo{year}{2022}.
\newblock \bibinfo{title}{Efficient sampling and structure learning of
  {B}ayesian networks}.
\newblock \bibinfo{journal}{Journal of Computational and Graphical Statistics}
  \bibinfo{volume}{31}, \bibinfo{pages}{639--650}.
\bibitem[{Leonelli(2019)}]{leonelli2019sensitivity}
\bibinfo{author}{Leonelli, M.}, \bibinfo{year}{2019}.
\newblock \bibinfo{title}{Sensitivity analysis beyond linearity}.
\newblock \bibinfo{journal}{International Journal of Approximate Reasoning}
  \bibinfo{volume}{113}, \bibinfo{pages}{106--118}.
\bibitem[{Leonelli et~al.(2020)Leonelli, Riccomagno and
  Smith}]{leonelli2020coherent}
\bibinfo{author}{Leonelli, M.}, \bibinfo{author}{Riccomagno, E.},
  \bibinfo{author}{Smith, J.Q.}, \bibinfo{year}{2020}.
\newblock \bibinfo{title}{Coherent combination of probabilistic outputs for
  group decision making: An algebraic approach}.
\newblock \bibinfo{journal}{OR Spectrum} \bibinfo{volume}{42},
  \bibinfo{pages}{499--528}.
\bibitem[{Leonelli and Varando(2022a)}]{leonelli2022highly}
\bibinfo{author}{Leonelli, M.}, \bibinfo{author}{Varando, G.},
  \bibinfo{year}{2022}a.
\newblock \bibinfo{title}{Highly efficient structural learning of sparse staged
  trees}, in: \bibinfo{booktitle}{International Conference on Probabilistic
  Graphical Models}, \bibinfo{organization}{PMLR}. pp.
  \bibinfo{pages}{193--204}.
\bibitem[{Leonelli and Varando(2022b)}]{leonelli2022structural}
\bibinfo{author}{Leonelli, M.}, \bibinfo{author}{Varando, G.},
  \bibinfo{year}{2022}b.
\newblock \bibinfo{title}{Structural learning of simple staged trees}.
\newblock \bibinfo{journal}{arXiv:2203.04390} .
\bibitem[{Leonelli and Varando(2023a)}]{leonelli2023context}
\bibinfo{author}{Leonelli, M.}, \bibinfo{author}{Varando, G.},
  \bibinfo{year}{2023}a.
\newblock \bibinfo{title}{Context-specific causal discovery for categorical
  data using staged trees}, in: \bibinfo{booktitle}{International Conference on
  Artificial Intelligence and Statistics}, pp. \bibinfo{pages}{8871--8888}.
\bibitem[{Leonelli and Varando(2023b)}]{leonelli2023learning}
\bibinfo{author}{Leonelli, M.}, \bibinfo{author}{Varando, G.},
  \bibinfo{year}{2023}b.
\newblock \bibinfo{title}{Learning and interpreting asymmetry-labeled {DAGs: A
  case study on COVID-19 fear}}.
\newblock \bibinfo{journal}{arXiv:2301.00629} .
\bibitem[{Liew et~al.(2022)Liew, de-la Llave-Rinc{\'o}n, Scutari,
  Arias-Bur{\'\i}a, Cook, Cleland and Fern{\'a}ndez-de
  Las-Pe{\~n}as}]{liew2022short}
\bibinfo{author}{Liew, B.X.}, \bibinfo{author}{de-la Llave-Rinc{\'o}n, A.I.},
  \bibinfo{author}{Scutari, M.}, \bibinfo{author}{Arias-Bur{\'\i}a, J.L.},
  \bibinfo{author}{Cook, C.E.}, \bibinfo{author}{Cleland, J.},
  \bibinfo{author}{Fern{\'a}ndez-de Las-Pe{\~n}as, C.}, \bibinfo{year}{2022}.
\newblock \bibinfo{title}{Do short-term effects predict long-term improvements
  in women who receive manual therapy or surgery for carpal tunnel syndrome?
  {A} {B}ayesian network analysis of a randomized clinical trial}.
\newblock \bibinfo{journal}{Physical Therapy} \bibinfo{volume}{102},
  \bibinfo{pages}{pzac015}.
\bibitem[{Mandhani et~al.(2020)Mandhani, Nayak and
  Parida}]{mandhani2020interrelationships}
\bibinfo{author}{Mandhani, J.}, \bibinfo{author}{Nayak, J.K.},
  \bibinfo{author}{Parida, M.}, \bibinfo{year}{2020}.
\newblock \bibinfo{title}{{Interrelationships among service quality factors of
  {M}etro {R}ail {T}ransit {S}ystem: An integrated {B}ayesian networks and
  {PLS}-{SEM} approach}}.
\newblock \bibinfo{journal}{Transportation Research Part A: Policy and
  Practice} \bibinfo{volume}{140}, \bibinfo{pages}{320--336}.
\bibitem[{Mandhani et~al.(2021)Mandhani, Nayak and
  Parida}]{mandhani2021establishing}
\bibinfo{author}{Mandhani, J.}, \bibinfo{author}{Nayak, J.K.},
  \bibinfo{author}{Parida, M.}, \bibinfo{year}{2021}.
\newblock \bibinfo{title}{Establishing service quality interrelations for
  {M}etro rail transit: {D}oes gender really matter?}
\newblock \bibinfo{journal}{Transportation Research Part D: Transport and
  Environment} \bibinfo{volume}{97}, \bibinfo{pages}{102888}.
\bibitem[{Marcot and Penman(2019)}]{marcot2019advances}
\bibinfo{author}{Marcot, B.G.}, \bibinfo{author}{Penman, T.D.},
  \bibinfo{year}{2019}.
\newblock \bibinfo{title}{Advances in {B}ayesian network modelling: Integration
  of modelling technologies}.
\newblock \bibinfo{journal}{Environmental Modelling \& Software}
  \bibinfo{volume}{111}, \bibinfo{pages}{386--393}.
\bibitem[{Pearl(1988)}]{pearl1988probabilistic}
\bibinfo{author}{Pearl, J.}, \bibinfo{year}{1988}.
\newblock \bibinfo{title}{Probabilistic reasoning in intelligent systems:
  Networks of plausible inference}.
\newblock \bibinfo{publisher}{Morgan Kaufmann}.
\bibitem[{Pearl(2009)}]{pearl2009causality}
\bibinfo{author}{Pearl, J.}, \bibinfo{year}{2009}.
\newblock \bibinfo{title}{Causality}.
\newblock \bibinfo{publisher}{Cambridge University Press}.
\bibitem[{Pensar et~al.(2015)Pensar, Nyman, Koski and
  Corander}]{pensar2015labeled}
\bibinfo{author}{Pensar, J.}, \bibinfo{author}{Nyman, H.},
  \bibinfo{author}{Koski, T.}, \bibinfo{author}{Corander, J.},
  \bibinfo{year}{2015}.
\newblock \bibinfo{title}{Labeled directed acyclic graphs: A generalization of
  context-specific independence in directed graphical models}.
\newblock \bibinfo{journal}{Data Mining and Knowledge Discovery}
  \bibinfo{volume}{29}, \bibinfo{pages}{503--533}.
\bibitem[{Pensar et~al.(2016)Pensar, Nyman, Lintusaari and
  Corander}]{pensar2016role}
\bibinfo{author}{Pensar, J.}, \bibinfo{author}{Nyman, H.},
  \bibinfo{author}{Lintusaari, J.}, \bibinfo{author}{Corander, J.},
  \bibinfo{year}{2016}.
\newblock \bibinfo{title}{The role of local partial independence in learning of
  {B}ayesian networks}.
\newblock \bibinfo{journal}{International Journal of Approximate Reasoning}
  \bibinfo{volume}{69}, \bibinfo{pages}{91--105}.
\bibitem[{Perucca and Salini(2014)}]{perucca2014travellers}
\bibinfo{author}{Perucca, G.}, \bibinfo{author}{Salini, S.},
  \bibinfo{year}{2014}.
\newblock \bibinfo{title}{{Travellers’ satisfaction with railway transport: A
  Bayesian network approach}}.
\newblock \bibinfo{journal}{Quality Technology \& Quantitative Management}
  \bibinfo{volume}{11}, \bibinfo{pages}{71--84}.
\bibitem[{Renooij(2001)}]{renooij2001probability}
\bibinfo{author}{Renooij, S.}, \bibinfo{year}{2001}.
\newblock \bibinfo{title}{Probability elicitation for belief networks: Issues
  to consider}.
\newblock \bibinfo{journal}{The Knowledge Engineering Review}
  \bibinfo{volume}{16}, \bibinfo{pages}{255--269}.
\bibitem[{Russell and Norvig(2009)}]{russel}
\bibinfo{author}{Russell, S.}, \bibinfo{author}{Norvig, P.},
  \bibinfo{year}{2009}.
\newblock \bibinfo{title}{Artificial Intelligence: A Modern Approach}.
\newblock \bibinfo{publisher}{Prentice Hall}.
\bibitem[{Salini and Kenett(2009)}]{salini2009bayesian}
\bibinfo{author}{Salini, S.}, \bibinfo{author}{Kenett, R.S.},
  \bibinfo{year}{2009}.
\newblock \bibinfo{title}{Bayesian networks of customer satisfaction survey
  data}.
\newblock \bibinfo{journal}{Journal of Applied Statistics}
  \bibinfo{volume}{36}, \bibinfo{pages}{1177--1189}.
\bibitem[{Scanagatta et~al.(2019)Scanagatta, Salmer{\'o}n and
  Stella}]{scanagatta2019survey}
\bibinfo{author}{Scanagatta, M.}, \bibinfo{author}{Salmer{\'o}n, A.},
  \bibinfo{author}{Stella, F.}, \bibinfo{year}{2019}.
\newblock \bibinfo{title}{A survey on {B}ayesian network structure learning
  from data}.
\newblock \bibinfo{journal}{Progress in Artificial Intelligence}
  \bibinfo{volume}{8}, \bibinfo{pages}{425--439}.
\bibitem[{Scutari(2010)}]{Scutari2010}
\bibinfo{author}{Scutari, M.}, \bibinfo{year}{2010}.
\newblock \bibinfo{title}{{Learning Bayesian networks with the bnlearn R
  package}}.
\newblock \bibinfo{journal}{Journal of Statistical Software}
  \bibinfo{volume}{35}, \bibinfo{pages}{1--22}.
\bibitem[{Scutari and Denis(2021)}]{denis}
\bibinfo{author}{Scutari, M.}, \bibinfo{author}{Denis, J.B.},
  \bibinfo{year}{2021}.
\newblock \bibinfo{title}{Bayesian networks: With examples in {R}}.
\newblock \bibinfo{publisher}{CRC press}.
\bibitem[{Scutari et~al.(2019)Scutari, Graafland and
  Guti{\'e}rrez}]{scutari2019learns}
\bibinfo{author}{Scutari, M.}, \bibinfo{author}{Graafland, C.E.},
  \bibinfo{author}{Guti{\'e}rrez, J.M.}, \bibinfo{year}{2019}.
\newblock \bibinfo{title}{Who learns better {B}ayesian network structures:
  Accuracy and speed of structure learning algorithms}.
\newblock \bibinfo{journal}{International Journal of Approximate Reasoning}
  \bibinfo{volume}{115}, \bibinfo{pages}{235--253}.
\bibitem[{Scutari and Nagarajan(2013)}]{scutari}
\bibinfo{author}{Scutari, M.}, \bibinfo{author}{Nagarajan, R.},
  \bibinfo{year}{2013}.
\newblock \bibinfo{title}{On identifying significant edges in graphical models
  of molecular networks}.
\newblock \bibinfo{journal}{Artificial Intelligence in Medicine}
  \bibinfo{volume}{57}, \bibinfo{pages}{207--217}.
\bibitem[{Silander and Leong(2013)}]{silander2013dynamic}
\bibinfo{author}{Silander, T.}, \bibinfo{author}{Leong, T.Y.},
  \bibinfo{year}{2013}.
\newblock \bibinfo{title}{A dynamic programming algorithm for learning chain
  event graphs}, in: \bibinfo{booktitle}{Proceedings of the 16th International
  Conference in Discovery Science}, pp. \bibinfo{pages}{201--216}.
\bibitem[{Smith and Anderson(2008)}]{smith2008conditional}
\bibinfo{author}{Smith, J.Q.}, \bibinfo{author}{Anderson, P.E.},
  \bibinfo{year}{2008}.
\newblock \bibinfo{title}{Conditional independence and chain event graphs}.
\newblock \bibinfo{journal}{Artificial Intelligence} \bibinfo{volume}{172},
  \bibinfo{pages}{42--68}.
\bibitem[{Spirtes et~al.(2001)Spirtes, Glymour and Scheines}]{spirtes}
\bibinfo{author}{Spirtes, P.}, \bibinfo{author}{Glymour, C.},
  \bibinfo{author}{Scheines, R.}, \bibinfo{year}{2001}.
\newblock \bibinfo{title}{Causation, Prediction, and Search}.
\newblock \bibinfo{publisher}{MIT Press}.
\bibitem[{Strong and Smith(2022)}]{strong2022bayesian}
\bibinfo{author}{Strong, P.}, \bibinfo{author}{Smith, J.Q.},
  \bibinfo{year}{2022}.
\newblock \bibinfo{title}{Bayesian model averaging of chain event graphs for
  robust explanatory modelling}, in: \bibinfo{booktitle}{International
  Conference on Probabilistic Graphical Models}, \bibinfo{organization}{PMLR}.
  pp. \bibinfo{pages}{61--72}.
\bibitem[{Talvitie et~al.(2019)Talvitie, Eggeling and
  Koivisto}]{talvitie2019learning}
\bibinfo{author}{Talvitie, T.}, \bibinfo{author}{Eggeling, R.},
  \bibinfo{author}{Koivisto, M.}, \bibinfo{year}{2019}.
\newblock \bibinfo{title}{Learning {B}ayesian networks with local structure,
  mixed variables, and exact algorithms}.
\newblock \bibinfo{journal}{International Journal of Approximate Reasoning}
  \bibinfo{volume}{115}, \bibinfo{pages}{69--95}.
\bibitem[{Thwaites et~al.(2010)Thwaites, Smith and
  Riccomagno}]{thwaites2010causal}
\bibinfo{author}{Thwaites, P.}, \bibinfo{author}{Smith, J.Q.},
  \bibinfo{author}{Riccomagno, E.}, \bibinfo{year}{2010}.
\newblock \bibinfo{title}{Causal analysis with chain event graphs}.
\newblock \bibinfo{journal}{Artificial Intelligence} \bibinfo{volume}{174},
  \bibinfo{pages}{889--909}.
\bibitem[{Thwaites et~al.(2008)Thwaites, Smith and
  Cowell}]{thwaites2008propagation}
\bibinfo{author}{Thwaites, P.A.}, \bibinfo{author}{Smith, J.Q.},
  \bibinfo{author}{Cowell, R.G.}, \bibinfo{year}{2008}.
\newblock \bibinfo{title}{Propagation using chain event graphs}, in:
  \bibinfo{booktitle}{Proceedings of the 24th Conference on Uncertainty in
  Artificial Intelligence}, pp. \bibinfo{pages}{546--553}.
\bibitem[{Tsamardinos et~al.(2006)Tsamardinos, Brown and
  Aliferis}]{tsamardinos2006max}
\bibinfo{author}{Tsamardinos, I.}, \bibinfo{author}{Brown, L.E.},
  \bibinfo{author}{Aliferis, C.F.}, \bibinfo{year}{2006}.
\newblock \bibinfo{title}{{The max-min hill-climbing Bayesian network structure
  learning algorithm}}.
\newblock \bibinfo{journal}{Machine Learning} \bibinfo{volume}{65},
  \bibinfo{pages}{31--78}.
\bibitem[{Varando et~al.(2021)Varando, Carli and Leonelli}]{varando2021staged}
\bibinfo{author}{Varando, G.}, \bibinfo{author}{Carli, F.},
  \bibinfo{author}{Leonelli, M.}, \bibinfo{year}{2021}.
\newblock \bibinfo{title}{Staged trees and asymmetry-labeled {DAGs}}.
\newblock \bibinfo{journal}{arXiv:2108.01994} .
\bibitem[{Viinikka and Koivisto(2020)}]{viinikka2020layering}
\bibinfo{author}{Viinikka, J.}, \bibinfo{author}{Koivisto, M.},
  \bibinfo{year}{2020}.
\newblock \bibinfo{title}{{Layering-MCMC for structure learning in Bayesian
  networks}}, in: \bibinfo{booktitle}{Proceedings of the 36th Conference on
  Uncertainty in Artificial Intelligence}, \bibinfo{organization}{PMLR}. pp.
  \bibinfo{pages}{839--848}.
\bibitem[{Wade(2023)}]{wade2023bayesian}
\bibinfo{author}{Wade, S.}, \bibinfo{year}{2023}.
\newblock \bibinfo{title}{Bayesian cluster analysis}.
\newblock \bibinfo{journal}{Philosophical Transactions of the Royal Society A}
  \bibinfo{volume}{381}, \bibinfo{pages}{20220149}.
\bibitem[{Walley et~al.(2023)Walley, Shenvi, Strong and
  Kobalczyk}]{walley2023cegpy}
\bibinfo{author}{Walley, G.}, \bibinfo{author}{Shenvi, A.},
  \bibinfo{author}{Strong, P.}, \bibinfo{author}{Kobalczyk, K.},
  \bibinfo{year}{2023}.
\newblock \bibinfo{title}{cegpy: Modelling with chain event graphs in
  {P}ython}.
\newblock \bibinfo{journal}{Knowledge-Based Systems} \bibinfo{volume}{274},
  \bibinfo{pages}{110615}.
\bibitem[{Werner et~al.(2017)Werner, Bedford, Cooke, Hanea and
  Morales-Napoles}]{werner2017expert}
\bibinfo{author}{Werner, C.}, \bibinfo{author}{Bedford, T.},
  \bibinfo{author}{Cooke, R.M.}, \bibinfo{author}{Hanea, A.M.},
  \bibinfo{author}{Morales-Napoles, O.}, \bibinfo{year}{2017}.
\newblock \bibinfo{title}{Expert judgement for dependence in probabilistic
  modelling: A systematic literature review and future research directions}.
\newblock \bibinfo{journal}{European Journal of Operational Research}
  \bibinfo{volume}{258}, \bibinfo{pages}{801--819}.
\bibitem[{Wilkerson and Smith(2021)}]{wilkerson2021customized}
\bibinfo{author}{Wilkerson, R.L.}, \bibinfo{author}{Smith, J.Q.},
  \bibinfo{year}{2021}.
\newblock \bibinfo{title}{Customized structural elicitation}.
\newblock \bibinfo{journal}{Expert Judgement in Risk and Decision Analysis} ,
  \bibinfo{pages}{83--113}.
\bibitem[{Xu et~al.(2020)Xu, Lu, Wang, Li and Zhang}]{xu2020improving}
\bibinfo{author}{Xu, X.}, \bibinfo{author}{Lu, Y.}, \bibinfo{author}{Wang, Y.},
  \bibinfo{author}{Li, J.}, \bibinfo{author}{Zhang, H.}, \bibinfo{year}{2020}.
\newblock \bibinfo{title}{Improving service quality of metro systems—{A} case
  study in the {B}eijing metro}.
\newblock \bibinfo{journal}{IEEE Access} \bibinfo{volume}{8},
  \bibinfo{pages}{12573--12591}.
\bibitem[{Yang et~al.(2022)Yang, Wang, Cheng and Chen}]{yang2022exploring}
\bibinfo{author}{Yang, M.}, \bibinfo{author}{Wang, Z.}, \bibinfo{author}{Cheng,
  L.}, \bibinfo{author}{Chen, E.}, \bibinfo{year}{2022}.
\newblock \bibinfo{title}{Exploring satisfaction with {air-HSR intermodal
  services: A Bayesian} network analysis}.
\newblock \bibinfo{journal}{Transportation Research Part A: Policy and
  Practice} \bibinfo{volume}{156}, \bibinfo{pages}{69--89}.
\bibitem[{Zhang and Thai(2016)}]{zhang2016expert}
\bibinfo{author}{Zhang, G.}, \bibinfo{author}{Thai, V.V.},
  \bibinfo{year}{2016}.
\newblock \bibinfo{title}{Expert elicitation and {B}ayesian network modeling
  for shipping accidents: A literature review}.
\newblock \bibinfo{journal}{Safety Science} \bibinfo{volume}{87},
  \bibinfo{pages}{53--62}.

\end{thebibliography}

\appendix

\section{Algorithm to learn k-parents staged trees}

\begin{algorithm}
    \SetKwInOut{Input}{Input}
    \SetKwInOut{Output}{Output}
    \SetKwComment{Comment}{/* }{ */}
\caption{Learning algorithm for k-parent staged trees using CMI}\label{alg:two}
\Input{A dataset $\mathcal{D}$ over categorical variables $X_1,\dots,X_p$ and $k\in\mathbb{Z}_+$}
\Output{A staged tree $T$}
\For{$i\gets1$ \KwTo $p$}{ 
$\Pi_i \gets \emptyset$\;
\eIf{$i\leq k+1$}{
$\Pi_i \gets [i-1]$\;
}{
\For{$j\gets 1$ \KwTo $k$}{
$max \gets -\infty$\;
\For{$s\in [i-1]\setminus\Pi_i$}{
\If{$I(X_i,X_j|X_{\Pi_i})>max$}{
$new \gets s$\;
}
}
$\Pi_i \gets \Pi_i \cup \{s\}$\;
}
}
}
Construct $G$ using $[p]$ and $\Pi_1,\dots,\Pi_p$\;
Transform $G$ to its equivalent staged tree $T$ with staging $U_1,\dots,U_{p-1}$\;
$score \gets BIC(T)$; $T^{*}\gets T$\;
\For{$i\gets 1$ \KwTo $p-1$}{
$indicator \gets 1$\; 
\While{$indicator\neq 0$}{
\For{every pair of stages $u_j,u_s\in U_i$}{
construct $T^{'}$ by merging $u_i$ and $u_j$\;
\If{$BIC(T^{'})<BIC(T^{*})$}{$score \gets BIC(T^{'})$; $T^{*} \gets T^{'}$\;}
}
\eIf{$T=T^{*}$}{$indicator \gets 0$}{$T\gets T^{*}$}
}
}
\Return $T$
\end{algorithm}

\end{document}